\RequirePackage[2020-02-02]{latexrelease}
\documentclass[pageno]{jpaper}

\usepackage[normalem]{ulem}
\usepackage{listings}
\usepackage{amsmath}
\usepackage[noend,linesnumbered,ruled,vlined]{algorithm2e}
\usepackage{physics}
\usepackage{amsmath}
\usepackage{multirow}
\usepackage{enumitem}
\usepackage{amssymb}
\newcommand{\ignore}[1]{}
\usepackage{adjustbox}
\usepackage[normalem]{ulem}
\usepackage[flushleft]{threeparttable}
\usepackage{dblfloatfix}

\usepackage{graphicx}
\usepackage{subfig}

\SetCommentSty{mycommfont}

\def\BibTeX{{\rm B\kern-.05em{\sc i\kern-.025em b}\kern-.08em
    T\kern-.1667em\lower.7ex\hbox{E}\kern-.125emX}}

\usepackage{url}

\usepackage[noadjust]{cite}
\usepackage{tikz}
\newcommand*\circled[1]{\tikz[baseline=(char.base)]{
            \node[shape=circle,draw,inner sep=0.5pt] (char) {#1};}}

\def\BibTeX{{\rm B\kern-.05em{\sc i\kern-.025em b}\kern-.08em
    T\kern-.1667em\lower.7ex\hbox{E}\kern-.125emX}}
\definecolor{aliceblue}{rgb}{0.94, 0.97, 1.0}
\usepackage{xcolor}
\usepackage{tcolorbox}
\tcbset{width=1\columnwidth,boxrule=1pt,colback=aliceblue,arc=1pt,left=2pt,right=2pt,boxsep=0.5pt}

\usepackage{authblk}
\begin{document}

\author{Poulami Das}
\author{Suhas K. Vittal}
\author{Moinuddin Qureshi}
\affil{Georgia Institute of Technology}
\affil{ \{poulami, suhaskvittal, moin\}@gatech.edu}
\title{
ForeSight: Reducing SWAPs in NISQ Programs \\ via Adaptive Multi-Candidate Evaluations}

\date{}
\maketitle

\thispagestyle{empty}

\begin{abstract}
Near-term quantum computers are noisy and have limited connectivity between qubits. Compilers are required to introduce SWAP operations in order to perform two-qubit gates between non-adjacent qubits. SWAPs increase the number of gates and depth of programs, making them even more vulnerable to errors. Moreover, they relocate qubits which affect SWAP selections for future gates in a program. Thus, compilers must select SWAP routes that not only minimize the overheads for the current operation, but also for future gates. Existing compilers tend to select paths with the fewest SWAPs for the current operations, but do not evaluate the impact of the relocations from the selected SWAP candidate on future SWAPs.
Also, they converge on SWAP candidates for the current operation and only then decide SWAP routes for future gates, thus severely restricting the SWAP candidate search space for future operations.

We propose {\em ForeSight}, a compiler that simultaneously evaluates multiple SWAP candidates for several operations into the future, delays SWAP selections to analyze their impact on future SWAP decisions and avoids early convergence on sub-optimal candidates. Moreover, ForeSight  evaluates slightly longer SWAP routes for  current operations if they have the potential to reduce SWAPs for future gates, thus reducing SWAPs for the program globally. As compilation proceeds, ForeSight dynamically adds new SWAP candidates to the solution space and eliminates the weaker ones. This allows ForeSight to reduce SWAP overheads at program-level while keeping the compilation complexity tractable. Our evaluations with a hundred benchmarks across three devices show that ForeSight reduces SWAP overheads by 17\% on average and 81\% in the best-case, compared to the baseline. ForeSight takes minutes, making it scalable to large programs.

\end{abstract}

\section{Introduction}
Quantum computers with only fifty-plus qubits can already outperform supercomputers for certain artificial tasks~\cite{QCSup,chinesesupremacy,ibmeagle}. These systems {\em Noisy Intermediate Scale Quantum (NISQ)}~\cite{preskillNISQ} are projected to scale up-to a thousand qubits by 2023~\cite{ibmq1000qubitroadmap} and are expected to power industry-scale applications soon~\cite{googleQcloud,qaoa,vqe}. Unfortunately, qubit devices are noisy and their high error-rates limit the fidelity of applications. For example, the average error-rate of a two-qubit operation is 1\% on existing quantum hardware~\cite{IBMQ,sycamoredatasheet}. Although increasing system sizes represents a significant milestone, NISQ systems would still be too small to achieve fault-tolerance. Instead, they will be operated in the presence of noise. Thus, improving the application fidelity by mitigating the impact of hardware errors is crucial to run real-world problems on NISQ devices~\cite{FCNature,CCS}.

\begin{figure*}[htb]
\centering
  \includegraphics[width=\linewidth]{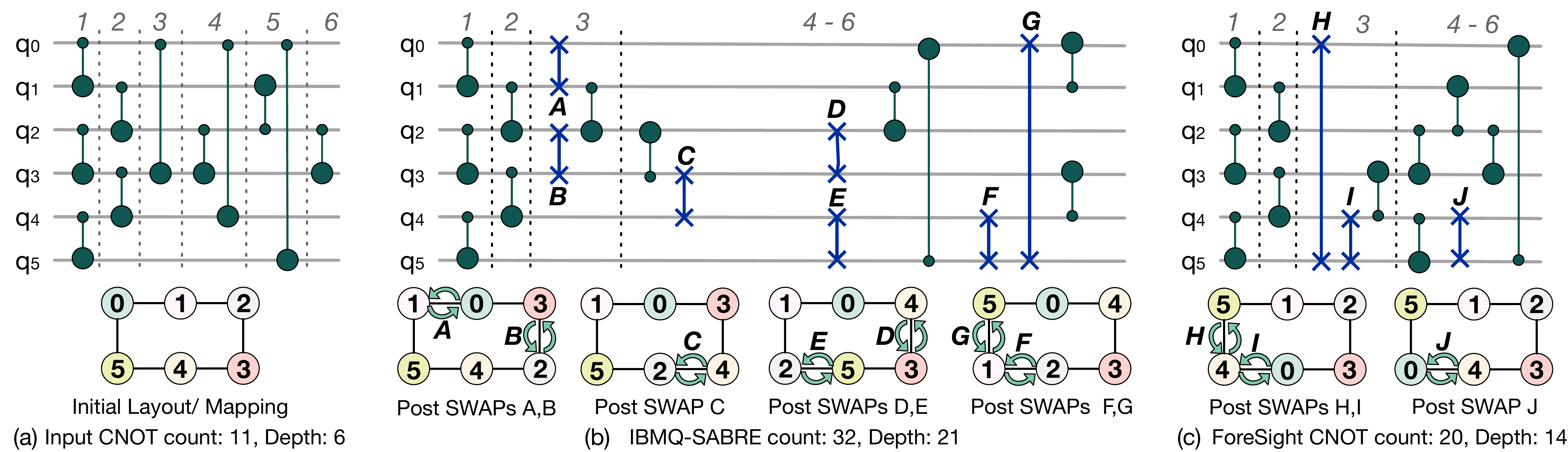}
\caption{(a) A circuit with six layers and initial layout. Compilation with intermediate steps by (b) IBMQ-SABRE and (c) ForeSight. IBMQ-SABRE minimizes SWAP overheads in each layer, whereas ForeSight reduces it globally.}
\label{fig:introfigure}
\end{figure*}

Besides high error-rates, most NISQ machines have limited connectivity between qubits due to fundamental device-level challenges~\cite{IBMQ,sycamoredatasheet}. On the other hand, two-qubit CNOT gates can only be performed on physically connected qubits. To perform a CNOT between non-adjacent qubits, compilers insert {\em SWAP} operations. A {\em SWAP} exchanges the state of two qubits and  can relocate any pair of non-adjacent qubits to physically connected locations. This process is called \textit{qubit movement} or \textit{qubit routing} and the number of SWAPs required depends on the device topology and distance between the non-adjacent qubits. Each SWAP consists of three serial CNOTs and therefore, increases the overall gate count and circuit depth. Thus, while SWAPs enable us to overcome the restricted connectivity of NISQ devices, they further increase the vulnerability of programs to hardware errors. To reduce the impact of SWAPs on application fidelity, compilers tend to choose qubit routing paths with minimum SWAPs~\cite{li2018tackling,zulehner2018efficient}. Unfortunately, SWAP minimization is an NP-complete problem~\cite{siraichi2018qubit} and despite several optimizations, SWAPs incur huge overheads and limit us from running large programs~\cite{harrigan2021quantum}. 

To compile a program, existing compilers (1)~decompose the circuit into layers, (2)~schedule the gates without any unresolved data dependencies or connectivity constraints, (3)~use greedy heuristics to insert minimal SWAPs for any CNOT that cannot be performed before moving to next layers, and (4)~terminate when all layers are processed. However, SWAPs not only increase the number of gates and circuit depth in the current layer, but also relocate qubits relative to each other that affect scheduling and SWAP selections for future gates. If not selected carefully, a current SWAP may introduce a greater number of SWAPs in future. Thus, compilers must select SWAPs that (1)~minimize gate overheads and depth for current operations; and  (2)~minimize SWAPs for future gates. 

\begin{tcolorbox}
\textbf{Takeaway 1}: Current compilers tend to choose SWAP routes with  minimum SWAPs, even if a longer SWAP route at present may result in fewer SWAPs in future. \\
\textbf{Takeaway 2}:
While deciding a SWAP candidate, compilers must optimize across two dimensions for future gates: (1)~maximize the number of SWAP-free gates that can be directly scheduled and (2)~minimize the number of SWAPs for future gates (thus, reducing SWAPs globally). Existing compilers with lookahead capabilities only optimize for maximum SWAP-free gates and not the other constraint. Moreover, regardless of the lookahead distance, they converge on the SWAP route for the current operation, and all future searches are done within this restricted search space. 
\end{tcolorbox}

We describe the limitations of existing compilers using an example. Figure~\ref{fig:introfigure}(a) shows a circuit and a NISQ device topology. IBM's Qiskit using the SABRE algorithm~\cite{li2018tackling}, maps the program qubits to the physical qubits and breaks the circuit into six layers, as shown in Figure~\ref{fig:introfigure}(a). Compilation proceeds from the lowest layer to the highest (to capture data dependencies). Gates are scheduled if feasible (CNOTs in Layers 1 and 2 for instance), whereas SWAPs are added when a CNOT that cannot be directly performed is encountered. For example, the CNOT in Layer 3 requires SWAPs as qubits $q_0$ and $q_3$ are non-adjacent. IBMQ-SABRE greedily selects SWAPs $A$ and $B$ to maximize parallelism and minimize the gate overheads, as shown in Figure~\ref{fig:introfigure}(b). As part of its lookahead feature, IBMQ-SABRE selects these SWAPs because they also enable the CNOT in Layer 4 without requiring any extra qubit movements (thus, maximizing the number of SWAP-free gates in the future). However, the CNOT in Layer 5 requires SWAPs post the relocations from SWAPs $A$ and $B$, but IBMQ-SABRE does not account for this factor while determining the SWAPs for Layer 3. IBMQ-SABRE proceeds to the next layers \textit{only after} the SWAPs for Layer 3 have been finalized and the second set of SWAPs ($C$, $D$, and $E$) are decided only upon reaching Layer 5. Figure~\ref{fig:introfigure}(b) shows the IBMQ-SABRE schedule with seven SWAPs. Note that while we discuss the limitations for IBMQ-SABRE, similar limitations exist for other compilers.

We propose {\em ForeSight}, that overcomes the drawbacks of existing compilers by (1) relaxing the constraints on the greedy heuristics and (2) delaying the decisions for SWAP candidate selections. Relaxing the constraints enables ForeSight to account for the temporal and spatial locality of the program qubits and explore a richer set of SWAP candidates that may be locally sub-optimal, but eventually brings together qubits on which majority of the operations are performed in future layers. Thus, the local cost is amortized over the scheduling of future operations. For example, as shown in Figure~\ref{fig:introfigure}(c), ForeSight relaxes the constraint of minimal CNOT and depth on qubit $q_0$ while evaluating SWAP candidates for the CNOT in Layer 3. It eventually selects a longer SWAP route with two SWAPs, $H$ and $I$, that displaces qubit $q_0$ by two locations and increases the depth compared to SWAPs $A$ and $B$ from IBMQ-SABRE. However, this cost is amortized as scheduling of gates in the future layers require fewer SWAPs. Overall, ForeSight uses only three SWAPs, as shown in Figure~\ref{fig:introfigure}(c).

The other drawback of current compilers is that they finalize SWAP candidates for the current layer before proceeding further and do not analyze the role of a current SWAP on future SWAP decisions. To tackle this challenge, ForeSight delays decisions about SWAP candidate selection and evaluates multiple SWAP candidates for several layers into the future. This allows ForeSight to evaluate the impact of a current SWAP in reducing SWAPs in future layers. As the cost of such exploration scales exponentially with the program size, ForeSight \textit{continuously prunes} the solution space  by discarding low-quality solutions while simultaneously adding newer SWAP candidates as the compilation progresses to keep only up to a maximum number of allowed candidates.

We observe that ForeSight and IBMQ-SABRE can be combined for greater benefits to exploit the diversity in their gate scheduling and SWAP insertion policies. The resultant design, {\em Hybrid-ForeSight (ForeSight-H)}, lowers SWAP overheads compared to either compiler standalone. Our evaluations using a hundred benchmarks and three device topologies show that ForeSight reduces SWAP overheads by 17\% on average and by up-to 81\% in the best-case, compared to the baseline. ForeSight compiles within minutes, making it scalable to large machines and programs with hundreds of qubits.

Compilers can further minimize the impact of SWAPs on application fidelity by accounting for the device error-rates to choose better-than-worst-case SWAPs~\cite{noiseadaptive,tannu2019not}. ForeSight is compatible with such noise-adaptive policies and our evaluations show that it can improve fidelity by 1.2x on average and by up-to 1.4x, compared to a noise-agnostic implementation. 

\vspace{0.05in}
Overall, this paper makes the following contributions: 

\begin{enumerate}[leftmargin=0cm,itemindent=.5cm,labelwidth=\itemindent,labelsep=0cm,align=left, itemsep=0.05 cm, listparindent=0.5cm]

    \item We show that greedy heuristics and early convergence on SWAP candidates can be sub-optimal at application-level. 

    \item We propose {\em ForeSight} that relaxes the SWAP and depth minimization constraints locally while selecting SWAP candidates and evaluates these individual candidates simultaneously across multiple layers before converging on a solution.
    
    \item We propose {\em Continuous Pruning} that allows ForeSight to keep compilation tractable by continuously adapting the SWAP candidate space by removing weaker candidates and adding newer ones as compilation proceeds. 
    
    \item We introduce {\em Hybrid-ForeSight} that combines ForeSight with existing compilers to further reduce SWAP overheads.

\end{enumerate}

\section{Background and Motivation}

\subsection{Impact of Limited Connectivity on Fidelity}
\begin{figure}[htp]
\centering
    \includegraphics[width=\columnwidth]{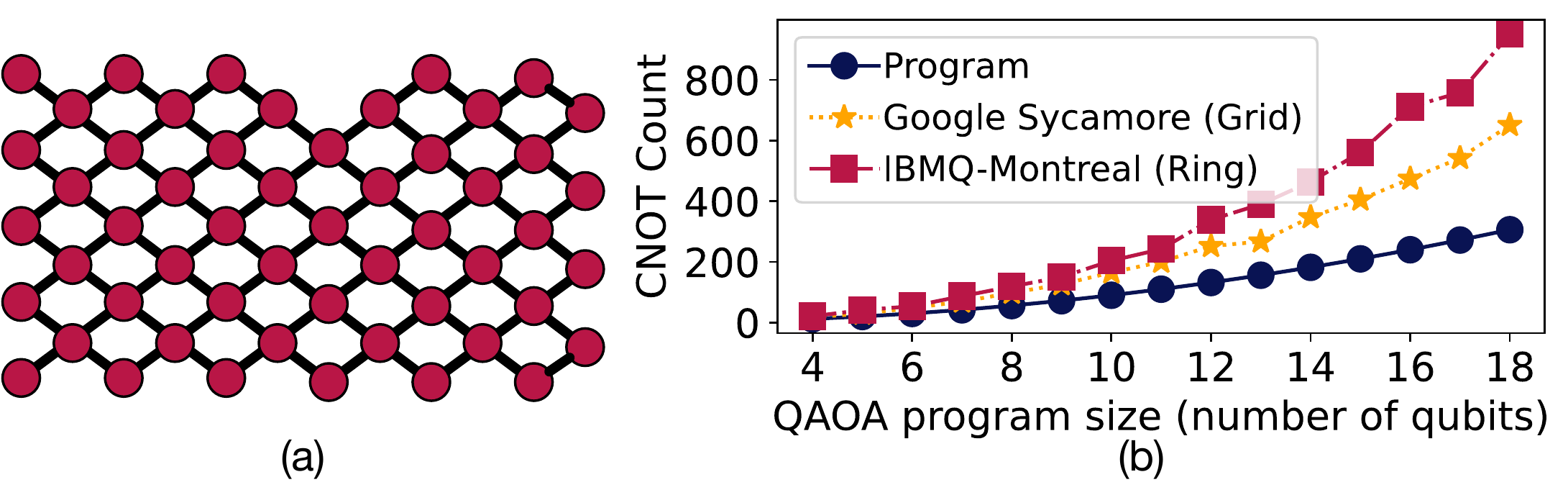}\vspace{-0.1in}
    \caption{(a) Google Sycamore device has grid topology where qubits are connected to some of their nearest neighbors. (b) Impact of SWAPs on the total CNOT count for QAOA benchmarks corresponding to MaxCut problems on fully connected graphs or the Sherrington Kirkpatrick (SK) model~\cite{sherrington1975solvable} for two device topologies (grid and ring).}
    \label{fig:swapoverheads}
\end{figure}

Most existing quantum computers have limited connectivity between qubits. Physically connecting two qubits on the device requires resonators operating at a dedicated frequency. The number of resonators required for all-to-all connectivity scales exponentially with the system size and therefore, becomes impractical. Also, too many connections on a single qubit deteriorate the fidelities of the gate operations due to crosstalk~\cite{hertzberg2021laser,zhang2020high}. For example, Figure~\ref{fig:swapoverheads}(a) shows the device topology of the state-of-the-art Google Sycamore~\cite{sycamoredatasheet}. 

To overcome the limited connectivity of NISQ devices, compilers insert {\em SWAPs} upon encountering any 2-qubit gates between non-adjacent qubits. A SWAP is a sequence of three CNOTs that exchanges the state of two qubits at the cost of increased gate count and circuit depth. The number of SWAPs required to relocate two qubits depends on the distance between them and device topology. Also, SWAPs for any current gate displace qubits that impact SWAPs needed in future. Ignoring this effect increases SWAP overheads even further if a current SWAP introduces more SWAPs in future and the problem scales with program size. Figure~\ref{fig:swapoverheads}(b) shows the number of CNOTs post SWAP insertion for QAOA MaxCut problems~\cite{qaoa} on fully connected graphs or the Sherrington Kirkpatrick (SK) model~\cite{sherrington1975solvable} for two different device topologies. SWAPs limit the fidelity of these programs beyond seventeen qubits on the Google Sycamore device~\cite{harrigan2021quantum}.

\subsection{Related Work on SWAP Reduction}
Compilers select qubit movement paths with fewest SWAPs and maximum parallelism as it increases the probability of successfully executing a program. Unfortunately, SWAP minimization is an NP-Complete problem~\cite{siraichi2018qubit}. Some NISQ compilers use solvers~\cite{bhattacharjee2017depth,booth2018comparing,lye2015determining,oddi2018greedy,shafaei2013optimization,shafaei2014qubit,venturelli2017temporal,venturelli2018compiling,wille2014optimal,nannicini2021optimal}, graph-partitioning~\cite{chakrabarti2011linear}, and dynamic programming~\cite{siraichi2018qubit}. However, these approaches incur long latencies that make them hard to scale and limit their adoption for practical applications with hundreds of qubits and quantum operations. More recently, industry-standard tool-chains have relied on heuristic compilers~\cite{QISKit,qmap,smith2020open,googletketcompiler}. For example, the heuristic approach by Zulhener et al. decomposes a circuit into layers and uses A* search to find SWAP routes with minimum SWAPs for each CNOT in a layer that requires them~\cite{zulehner2018efficient}. SABRE, adopted in IBM's Qiskit tool-chain, is another heuristic compiler that determines SWAPs faster by only searching locally~\cite{li2018tackling}. However, these policies do not evaluate the impact of a SWAP route on future SWAP decisions, as explained next.

\subsection{Limitations of Current State-of-the-Art}
Existing compilers suffer from two major drawbacks:
\vspace{0.05in}
\begin{enumerate}[leftmargin=0cm,itemindent=.5cm,labelwidth=\itemindent,labelsep=0cm,align=left, itemsep=0.1 cm, listparindent=0.5cm]
    \item \textbf{Relying on greedy heuristics:} Current solutions tend to only select minimal SWAPs even if a costlier or longer SWAP route for a current gate could reduce SWAPs in future.
    \item \textbf{Early SWAP decisions limit lookahead capacity:} SWAPs displace qubits and affect scheduling and SWAP selections for future operations. While considering future gates during current SWAP selection, a compiler must optimize across two dimensions: (1)~maximizing the number of SWAP-free gates (that can be directly scheduled post the relocations caused by the current SWAP candidate) and (2)~minimizing the number of future SWAPs for those gates that require them. Unfortunately, existing compilation approaches do not account for the second dimension during SWAP candidate selection. 
\end{enumerate}

\ignore{
\vspace{0.05in}
\noindent \textbf{Why are current lookahead heuristics inadequate?} While current compilers use lookahead heuristics, they are limited to selecting SWAP candidates that only maximizes the number of SWAP-free gates in future. They consider two lists of gates: current and future, and select SWAPs that can schedule the maximum number of gates from both sets, with higher priority given to the current list. However, they do not assess the impact of qubit relocations from a current SWAP candidate on the selection of future SWAPs that will be required to perform gates that cannot be directly scheduled. For example, Figure~\ref{fig:basiclookahead}(a-b) shows a circuit and qubit layout. To perform CNOT \circled{0}, a compiler has two SWAP options, as shown in Figure~\ref{fig:basiclookahead}(c).  Here, the current list contains CNOT \circled{0}, whereas the future list contains CNOTs \circled{1},\circled{2}, and \circled{3}. While both routing options cost one SWAP, the relocations from Option-2 allow direct scheduling of CNOTs \circled{1} and \circled{2} from future and is selected. However, existing compilers do not evaluate the impact both these routing options on the SWAPs that will be required to perform CNOT \circled{3}.
Increasing the size of the future list cannot address this limitation. Moreover, existing compilers immediately converge on a SWAP candidate which severely restricts the decision space for future operations, as the current SWAPs have already been finalized. For example, the compilers select Option-2 and discard Option-1, even if Option-1 could have led to fewer SWAPs globally. 

\begin{figure}[htb] 
\includegraphics[width=\columnwidth]{./Figures/lookahead.pdf}
\caption{(a) A quantum circuit and (b) an initial layout. (c) Routing options that both require one SWAP, but Option-2 is selected as it also enables CNOTs 1 and 2.}
\label{fig:basiclookahead}
\end{figure} 
}

\begin{figure*}[htb]
\centering
  \includegraphics[width=\linewidth]{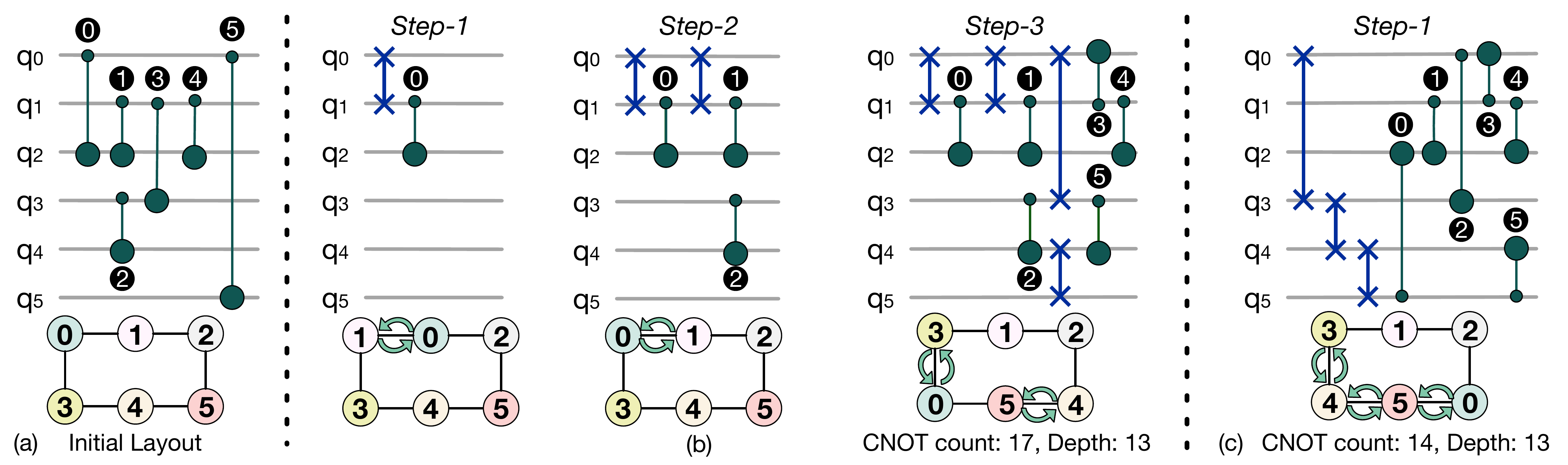}
\caption{(a) A quantum circuit and initial layout. (b) Circuit compilation by IBMQ-SABRE and intermediate layout after each round of SWAP insertion. In each step, IBMQ-SABRE greedily uses minimum SWAPs required to accomplish the CNOTs and maximizes the parallelism (Step-3 for example). (c) ForeSight relaxes the constraint of minimal SWAPs and depth on qubit $q_0$ and uses three SWAPs in Step-1 (compared to a single SWAP in Step-1 of IBMQ-SABRE). This cost is amortized over time because no more SWAPs are required in future layers. (Figure is for illustration only)}
\label{fig:constraintrelaxation}
\end{figure*}
\begin{tcolorbox}
Our \textit{goal} is to minimize SWAPs globally at the application-level by relaxing the constraints on greedy heuristics and delaying SWAP decisions to account for the impact of current selection on future SWAPs.
\end{tcolorbox}

\ignore{

\subsection{Goal: Minimize the Number of SWAPs}
SWAPs limit us from running large programs on NISQ devices. Our goal is to design a compiler that minimizes the overheads of SWAPs without significantly deteriorating the search complexity. To that end, we present {\em SWIM-- SWAP Insertion via Multi-path evaluations}. It leverages the following key insights to determine SWAP routes:

\begin{enumerate}[leftmargin=0cm,itemindent=.5cm,labelwidth=\itemindent,labelsep=0cm,align=left, itemsep=0.15cm, listparindent=0.3cm]

\item Instead of relying on greedy heuristics, relaxing the constraints on CNOT and depth minimization locally can yield better SWAP routes globally at the application-level. At a given phase of a program, if a particular qubit will not be used in the near future layers, relaxing the constraints on it during SWAP insertion can lead to more effective global optimization. 

\item Instead of inserting SWAPs one at a time, as required, in each layer before proceeding to the next layer, SWAP routes can be optimized further by evaluating multiple SWAP paths in parallel across several layers in the circuit together.

\item Instead of relying on a small and constant sized look-ahead window, SWAPs can be minimized further by dynamically adjusting the window size depending on the program characteristics in each layer during compilation.

\end{enumerate}

\vspace{0.05 in}

}

\section{ForeSight: Key Insights}
We present  {\em ForeSight}-- a compiler that introduces SWAPs by evaluating multiple candidate solutions. We discuss the key insights before describing the algorithm. 


\subsection{Relax Constraints during SWAP Insertion}\label{sec:foresight,ssec:insight1}

\noindent \textbf{Existing approaches}: For any CNOT that requires SWAPs, current compilers typically use greedy heuristics to insert SWAP routes one-at-a-time that result in minimum SWAPs and circuit depth. For example, Figure~\ref{fig:constraintrelaxation}(a) shows a program and layout. Figure~\ref{fig:constraintrelaxation}(b) shows the three SWAP insertion steps using IBMQ-SABRE. In the first two steps, only one SWAP is inserted, whereas two SWAPs are inserted in Step-3. The SWAP in Step-2 allows CNOTs \circled{1} and \circled{2} to be scheduled in parallel. Similarly, the SWAPs in Step-3 are executed in parallel which further allows parallel execution of CNOTs \circled{3} and \circled{4} next. Thus, these SWAP selections attempt to minimize the overheads of CNOTs and depth for each layer. 

\vspace{0.1in}

\noindent \textbf{Limitations}: Greedy SWAP insertion may be optimal locally, but can become sub-optimal at application-level as it does not account for program-specific characteristics. Always using greedy heuristics can result in poor qubit layout that includes a program qubit with no temporal locality to be included along the movement paths of qubits that will be frequently used in the next few operations. For example, qubit $q_0$ lies along the qubit movement paths of $q_1$, $q_2$, and $q_3$ which increases the overall SWAP cost, as shown in Figure~\ref{fig:constraintrelaxation}(b).

\vspace{0.1in}
\noindent \textbf{Insight- Relax constraints locally}: ForeSight relaxes the constraints on minimal SWAPs and depth locally, depending on the usage of program qubits. For example,  Figure~\ref{fig:constraintrelaxation}(c) shows that ForeSight inserts three SWAPs in Step-1 compared to one SWAP in IBMQ-SABRE. This is sub-optimal and a longer SWAP route if we only consider the possibilities for CNOT \circled{0}. But this cost is amortized while scheduling future operations. As $q_0$ is not used between CNOTs \circled{0} and \circled{5}, ForeSight exploits this temporal behavior to relocate this qubit farther away. While this movement is three times more expensive locally, it reduces the distance between the qubits on which CNOTs \circled{1} to \circled{4} are performed. Hence, ForeSight requires no further SWAPs for scheduling these CNOTs. 

\subsection{Concurrent Analysis of Multiple Solutions}
\label{sec:foresight,ssec:insight2}

\begin{figure*}[!htb]
\centering
  \includegraphics[width=\linewidth]{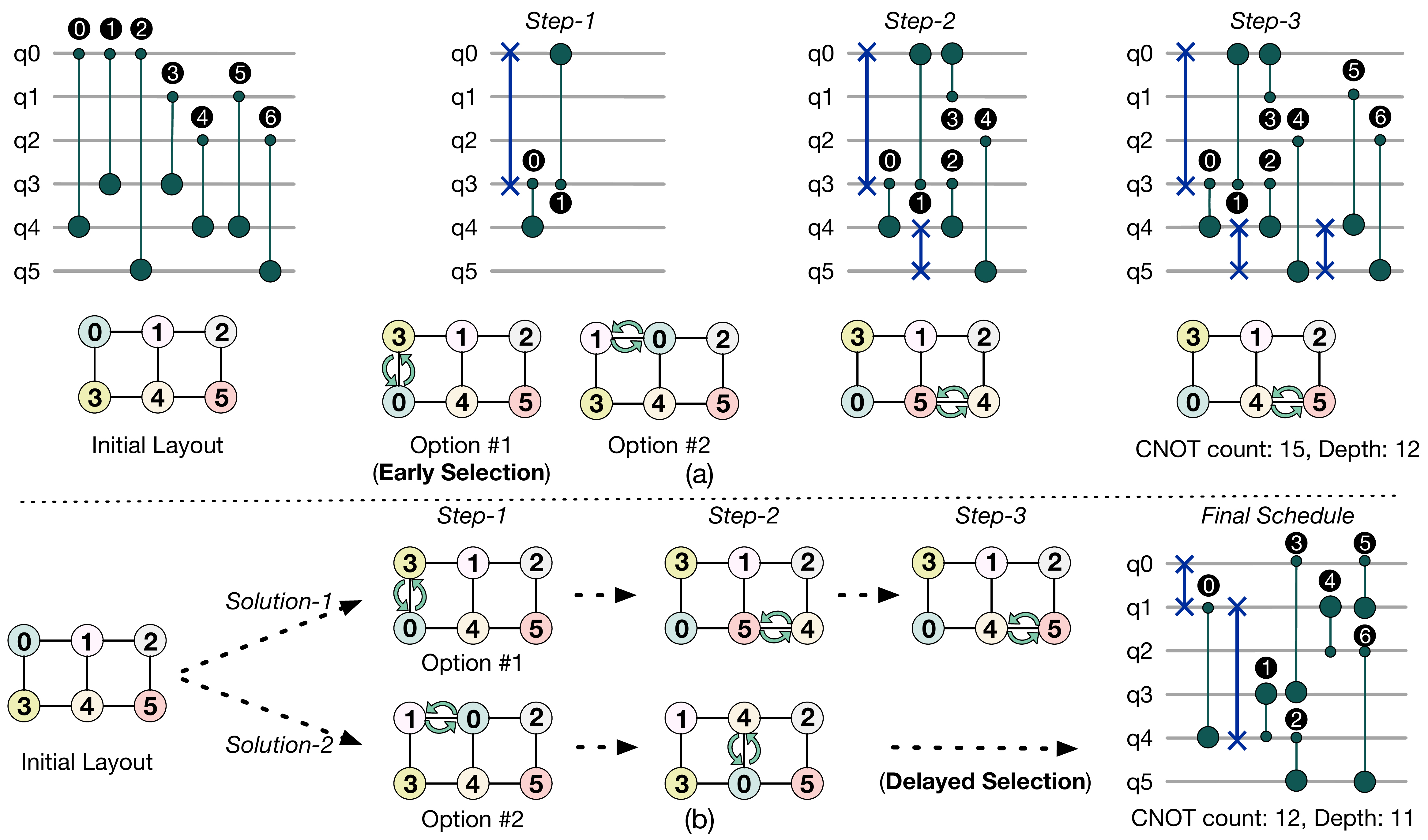}
\caption{(a) Example of a quantum circuit and an initial layout. Compilation by IBMQ-SABRE and the intermediate qubit layouts after each step of SWAP insertion. At Step-1, there are two potential SWAP candidates and Option-1 is retained as it allows depth minimization while scheduling CNOTs in Step-2. (b) Intermediate steps in ForeSight where multiple solutions are evaluated, and SWAP selection is delayed until Step-3. (Figure is for illustration purposes only)}
\label{fig:multipathevals}
\end{figure*}

\noindent \textbf{Existing approaches}: Current compilers select SWAPs for each CNOT in a layer that requires them before moving to the next layer. For example, Figure~\ref{fig:multipathevals} shows a  circuit and a layout. The compiler must insert SWAPs to execute CNOT \circled{0}. It has two options: (1) SWAP $q_0$, $q_3$ or (2) SWAP $q_0$, $q_1$, as shown in Figure~\ref{fig:multipathevals}(a). Both SWAPs incur equal CNOT overheads if existing lookahead features are considered as they both enable CNOT \circled{1} without any extra SWAPs, as shown in Step-1 of Figure~\ref{fig:multipathevals}(a). Existing compilers select SWAP $q_0$, $q_3$ (Option 1 over 2) as it also reduces depth by scheduling CNOT \circled{1} in parallel with the SWAP required for CNOT \circled{2}, as shown in Step-2 of Figure~\ref{fig:multipathevals}(a). Even worse, in reality, compilers randomly select a candidate before proceeding forward. 

\vspace{0.1in}
\noindent \textbf{Limitations}: Current compilers converge \textit{early} on SWAP candidates for each layer. This creates blind spots that may adversely lead to the selection of sub-optimal candidates. For example, the second option in Step-1 (discarded by existing compilers) leads to fewer SWAPs at the application-level, as shown in Figure~\ref{fig:multipathevals}(b). The probability of rejecting a good SWAP candidate early increases exponentially with the number of times the compiler makes a SWAP selection decision.

\vspace{0.1in}
\noindent \textbf{Insight- Delay SWAP route selections}: ForeSight overcomes this limitation by evaluating multiple SWAP candidates for many operations in the future, before converging on one of them. For example, when ForeSight finds two candidates with identical costs during the insertion of the SWAP in Step-1, it \textit{retains both of them}, unlike existing compilers which randomly pick one. Note that now the cost of both candidates is identical because the constraint on depth minimization is relaxed. As shown in Figure~\ref{fig:multipathevals}(b), ForeSight evaluates the impact of both SWAP candidates on the gates as well as SWAPs in the future layers. In Step-2, each path selects their individual SWAP candidates and continues scheduling (some candidates are not shown for simplicity of the illustration). As ForeSight proceeds, Path-1 requires an extra SWAP to accomplish CNOTs \circled{6} and \circled{7}, unlike Path-2, as shown in Step-3 of Figure~\ref{fig:multipathevals}(b). Finally, as ForeSight delays the SWAP selection decisions, it converges on the second solution that requires fewer SWAPs overall at the application-level.

\section{ForeSight: Design}
ForeSight compiles programs by (1)~relaxing SWAP insertion constraints locally to select a limited number of longer SWAP routes and (2)~delaying SWAP selections. ForeSight also accounts for device topology in its lookahead heuristics to capture its routing capability for future operations. In this section, we describe how the key insights are used to design ForeSight but summarize the notations used in the algorithm in Table~\ref{tab:notations} before discussing the overall implementation. 

\begin{table}[!htb]
\begin{center}
\begin{small}

\caption{Definition of notations used in ForeSight}
\setlength{\tabcolsep}{1.2mm} 
\renewcommand{\arraystretch}{1.0}
\label{tab:notations}
{
\begin{tabular}{ |c|l|} 
\hline
Notation & Definition \\
\hline
\hline
$n$ & Number of program qubits \\
\hline
$N$ & Number of physical qubits on the NISQ device. \\
\hline
$q_{\{1,2,\cdots,n\}}$ & Program qubits in the quantum circuit \\
\hline
$Q_{\{1,2,\cdots,N\}}$ & Physical qubits on the NISQ device \\
\hline
\multirow{2}{*}{$G(V, E)$} & Directed coupling graph of the NISQ device, \\
& where $V \in \{Q_1,...,Q_N\}$, $E= \{\textrm{Edges on } G \}$.\\
\hline
$\delta$ & Constraint relaxation factor \\
\hline
$P_{i\rightarrow j}$ & A path or route from $Q_i$ to $Q_j$\\
\hline
$P^{\textrm{min}}_{i\rightarrow j}$ & Shortest path from $Q_i$ to $Q_j$\\
\hline
$d()$ & Path length or distance between two physical qubits\\
\hline 
$D[$ $][$ $]$ & Distance matrix, $D[i][j]$ lists paths from $Q_i$ to $Q_j$\\
\hline
\ignore{
$\mathsf{Primary}$ & List of layers containing only two-qubit gates \\
\hline
$\mathsf{Secondary}$ & List of layers of single-qubit gates and barriers \\
\hline
}
$g$ & A list of program qubits on which a gate is applied\\
\hline
$F$ & Front layer of 2-qubit gates that requires routing\\

\hline
$\Delta$ & Relative distance between two DAG layers \\
\hline
$\mathsf{Post}$ & Array of gates in future layers of F \\
\hline
\multirow{2}{*}{$\pi( )$} & 1:1 mapping between program qubits $q_{\{1,2,\cdots,n\}}$ \\ 
& to physical qubits $Q_{\{1,2,\cdots,N\}}$\\
\hline
$\mu_G$ & Routing capacity or mean connectivity of $G$ \\
\hline
$H()$ & Heuristic cost function \\
\hline
$T$ &  A tree to evaluate multiple solutions concurrently \\
\hline
$\eta^i_j$ & Node in the $i^{th}$ layer of the solution tree $T$\\
\hline
$S$ & A solution of routed gates in the solution tree $T$\\
\hline
\end{tabular}}
\vspace{-0.2in}
\end{small}
\end{center}
\end{table}

\begin{figure*}[htb]
\centering
  \includegraphics[width=\linewidth]{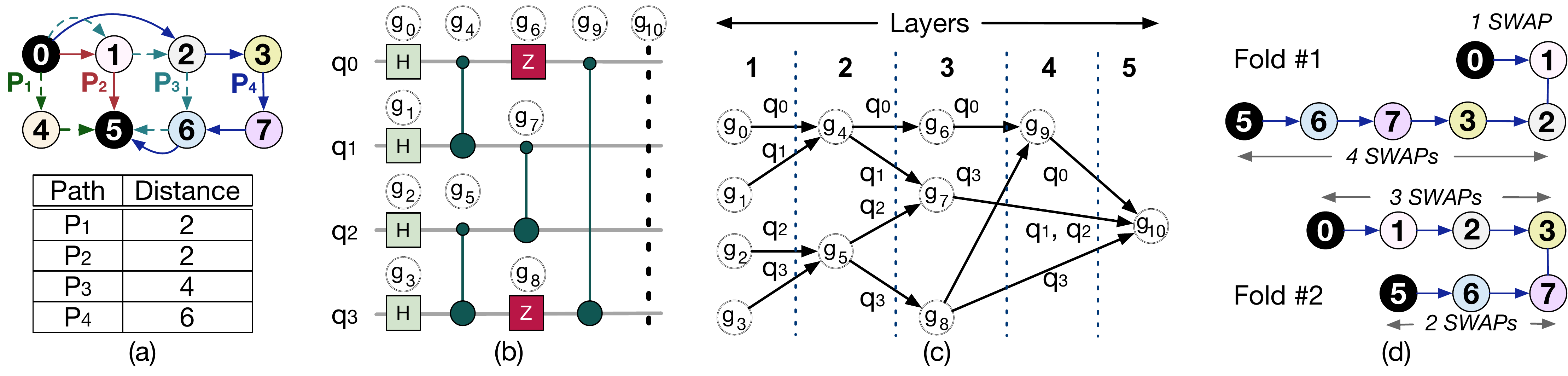}
\caption{(a) Path evaluations for computing Distance matrix $D[$ $][$ $]$. The shortest path between $Q_0$ and $Q_5$ is at a distance $2$. If the constraint relaxation factor is $2$, $D[0][5]$ stores $P_1$, $P_2$, and $P_3$ but not $P_4$. (b) A quantum circuit (c) translated into a Directed Acyclic Graph. (d) Example of \textit{foldings} for path $P_4$ that result in a critical path of 4 and 3 SWAPs respectively.}
\label{fig:algooverview}
\end{figure*}

\newpage
\subsection{Distance Matrix Computation}
We obtain the Distance matrix $D[$ $][$ $]$ of a device by analyzing its coupling graph $G(V, E)$ using the Floyd-Warshall algorithm~\cite{floyd1962algorithm}. Each entry $D[i][j]$ contains the list of paths between physical qubits $Q_i$ and $Q_j$, $[P_{i\rightarrow j}]$, such that the length of the path lies within the length of the shortest path between the qubits, $d(P^{\textrm{min}}_{i\rightarrow j})$, and $d(P^{\textrm{min}}_{i\rightarrow j})+\delta$, as described in Equation~\eqref{eq:paths}. Thus, each entry includes (i)~the shortest paths as well as (ii)~the paths that are within distance $\delta$ from the shortest path. For example, Figure~\ref{fig:algooverview}(a) shows the list of all paths between qubits $Q_0$ and $Q_5$ of which only three paths, $P_1$, $P_2$, and $P_3$, are stored in the distance matrix considering $\delta=2$, whereas $P_4$ is not stored. This enables us to select SWAP candidates by relaxing constraints, as will be explained in the next subsections. Note that this step is required only once.

\begin{equation}
    \label{eq:paths}
    D[i][j] = [P_{i\rightarrow j}], \forall \textrm{ } d(P^{\textrm{min}}_{i\rightarrow j})\leq d(P_{i\rightarrow j}) \leq d(P^{\textrm{min}}_{i\rightarrow j})+\delta
\end{equation}

\subsection{Circuit Decomposition into Layers}\label{sec:designimpl,ssec:circuitdecomp}
We convert the circuit into a Directed Acyclic Graph (DAG) to capture the data dependencies. Figure~\ref{fig:algooverview}(b-c) shows a circuit and its DAG. ForeSight maintains a list of gates that are yet to be executed as the front layer, $F$. In every iteration, ForeSight schedules those gates in $F$ whose qubits are connected on the device. If all gates in $F$ are scheduled, ForeSight sets $F$ to the next  layer. Alternately, ForeSight searches for SWAP candidates if it encounters at least one gate in $F$ that cannot be scheduled due to connectivity constraints.

\color{black}

\subsection{Shortlisting SWAP Candidates}\label{sec:designimpl,ssec:candselect}
ForeSight computes an array of two-qubit gates, $\mathsf{Post}$ using the layers after $F$. The entry $\mathsf{Post}[i]$ contains a list of tuples $(g,\Delta)$, where each tuple stores the gates using program qubit $q_i$ and their distance $\Delta$ from $F$. For example, $\mathsf{Post}[0]$ stores [($g_4$,1), ($g_7$,2)] because gates \circled{$g_4$} and \circled{$g_7$} are one and two layers from $F$ respectively. This captures the temporal locality and data dependencies of each qubit. ForeSight examines the next $\lceil 10\mu_G \rceil$ layers, where $\mu_G$ is the device routing capacity, and adds them to $\mathsf{Post}$. For a gate in $F$ on qubits $q_i$ and $q_j$, SWAPs are not needed if $\pi(q_i)$, $\pi(q_j)$ are connected and the gate is scheduled, where $\pi$ denotes a mapping between program to physical qubits. Alternately, ForeSight prepares a list of SWAP candidates if they are non-adjacent.

To select SWAP candidates, ForeSight looks up paths between $\pi(q_i)$ and $\pi(q_j)$ in the entry $D[\pi(q_i)][\pi(q_i)]$ and \textit{folds} them such that $q_i$ and $q_j$ meet at a location that minimizes our heuristic cost function. For example, Figure~\ref{fig:algooverview}(d) shows two possible folds for path $P_4$. The first option folds the path along edge $Q_1$-$Q_2$ such that the SWAPs relocate $Q_0$ to $Q_1$ and $Q_5$ to $Q_2$. This results in a critical path with 4 SWAPs. The second option folds the path along edge $Q_3$-$Q_7$ that relocates $Q_0$ to $Q_3$ and $Q_5$ to $Q_7$, resulting in a critical path with 3 SWAPs. By default, ForeSight folds a path along the edge that is midway between the end-points of a path to reduce the circuit depth and maximize parallelism. If $F$ contains multiple gates, ForeSight tries to merge the folded paths and returns all folded path combinations that allow scheduling every gate in $F$, maximizing parallelism. If ForeSight fails to find a combination that can satisfy all gates in $F$ (because the individual folded paths intersect), it splits $F$ into multiple layers with gates that comprise of non-intersecting folded paths and repeats the process until all gates in $F$ can be scheduled. Note that, unlike other compilers that introduce one SWAP at a time, ForeSight inserts a complete path of SWAPs. 

\subsection{Relaxing SWAP Selection Constraints}
\label{sec:designimpl,ssec:relax}
ForeSight relaxes SWAP selection constraints and evaluates a few locally sub-optimal candidates too. To implement this, ForeSight uses the \textit{constraint relaxation factor} $\delta$ while computing the distance matrix $D[$ $][$ $]$. The value $\delta=0$ corresponds when we consider only the shortest paths between two qubits, whereas $\delta=1$ enables ForeSight to select the shortest paths as well as paths that are at most one edge longer than the shortest paths. By default, ForeSight uses $\delta=2$. 

\subsection{Evaluating the Heuristic Cost Function}
ForeSight now contains \textit{\textbf{multiple SWAP candidates}} corresponding to $F$, including a few longer SWAP routes. We compute the cost of a SWAP candidate using a heuristic cost function, $H_{\textrm{total}}$, that combines the cost of the SWAP candidate ($H_{\textrm{SWAP candidate}}$) and the lookahead heuristic ($H_{\textrm{lookahead}}$). Retaining multiple SWAP candidates allows ForeSight to assess the impact of a SWAP candidate on future SWAP decisions and prevent early convergence on sub-optimal solutions.

\subsection{Adapting Lookahead for Device Topology}
The impact of a SWAP candidate on the routing of future gates depends on the program and the \textit{routing capacity} of the device. Devices with very sparse connectivity, like Rigetti Aspen, have low routing capacity because SWAP options are very limited. Alternately, devices with grid topology, such as Google Sycamore, have more routing paths per physical qubit and thus, ForeSight can relax constraints and find alternative routes more easily. We quantify the routing capacity, $\mu_G$, as the mean device connectivity, as shown in Equation~\eqref{eq:routingcapacity}. As most NISQ devices exhibit symmetry in the physical layout of qubits, using the global routing capacity is adequate, and region-specific local routing capacity is not required.
\begin{equation}
    \label{eq:routingcapacity}
    \mu_G = \frac{\textrm{Number of links on the device}}{\textrm{Number of physical qubits on the device}}
\end{equation}

To account for the device topology on future routing for a current SWAP candidate, we introduce $\mu_G$ in our lookahead heuristic, $H_{\textrm{lookahead}}$, as described in Equation~\eqref{eq:hlookahead}. This decaying heuristic ensures that the impact of the farthest gates in $\mathsf{Post}$ is minimal for devices with low routing capacity. We also account for the number of gates in $\mathsf{Post}$ so that 
ForeSight can maximally exploit the device connectivity. For instance, denser topologies can handle more gates in $\mathsf{Post}$ as they have more neighbors compared to sparse topologies.
\begin{equation}
    \label{eq:hlookahead}
    \resizebox{0.90\hsize}{!}{$
    H_\textrm{lookahead}=\frac{1}{\mid \mathsf{Post} \mid}\sum_{[g = [q_i,q_j],\Delta] \in \mathsf{Post}}d(\pi(q_i),\pi(q_j))\times e^{-{(\frac{\Delta}{\mu_G}})^2}$}
\end{equation}

 The total cost of a SWAP candidate, $H_{\textrm{total}}$, is calculated as the sum of the lookahead heuristic cost and the cost of an individual SWAP candidate in terms of the number of CNOTs scaled down per the size of $\mathsf{Post}$ and routing capacity $\mu_G$ of the device, as described in Equation~\eqref{eq:htotal}.
\begin{equation}
    \label{eq:htotal}
    H_\textrm{total}= H_\textrm{lookahead} + H_\textrm{SWAP candidate} \times e^{-{(\frac{\mid \mathsf{Post} \mid}{\mu_G}})^2}
\end{equation}

\ignore{
ForeSight creates a \textit{\textbf{SWAP candidate pool}} which comprises of a list of SWAP candidates for the given front layer, $F$, that minimizes the value of the heuristic cost function, $H_{\textrm{total}}$.
}

ForeSight creates a SWAP candidate pool which comprises of a list of SWAP candidates for the front layer $F$.

\subsection{Delaying SWAP Decisions}
ForeSight creates \textbf{\textit{multiple solutions}} independently originating from each candidate in the SWAP pool and constructs a \textit{solution tree}. For each SWAP candidate (as root), ForeSight continues scheduling by moving $F$ to the next layers of the DAG and repeats the SWAP selection process, as required, adding solutions or branches to the tree, as layers are processed. However, the tree formation does not remain tractable as its size grows quickly with increasing circuit size and depth. To tackle this challenge, ForeSight uses \textit{\textbf{continuous pruning}}. 
 
Continuous pruning ensures the tree size does not exceed the \textit{maximum number of solutions allowed} at any point. To enable this, ForeSight holds additional information in the tree. The tree consists of nodes and edges, where each node $\eta^i_j$ denotes a SWAP candidate $j$ for the circuit DAG layer $i$ and each edge represents a potential solution, $S$, of routed gates until layer $i$. A node $\eta^i_j$ in the $i^{\textrm{th}}$ layers stores (1)~a \textit{pointer} to its parent in the previous layer, $*\eta_{k}^{i-1}$, (2)~list of \textit{gates} to execute, $[g_j]$, and (3)~\textit{CNOT cost}. The CNOT cost of a node is the sum of the number of CNOTs required for the current node and its parent, as described in Equation~\eqref{eq:cnotsum}.\begin{equation}
\resizebox{0.90\hsize}{!}{$\begin{split}
\textrm{CNOT Cost for node }\eta^i_j =  \textrm{Number of CNOTs in } [g_j] + \\
\textrm{CNOT Cost for node }\eta^{i-1}_k
\end{split}$}\label{eq:cnotsum}
\end{equation}

\begin{figure}[htb]
  \includegraphics[width=\columnwidth]{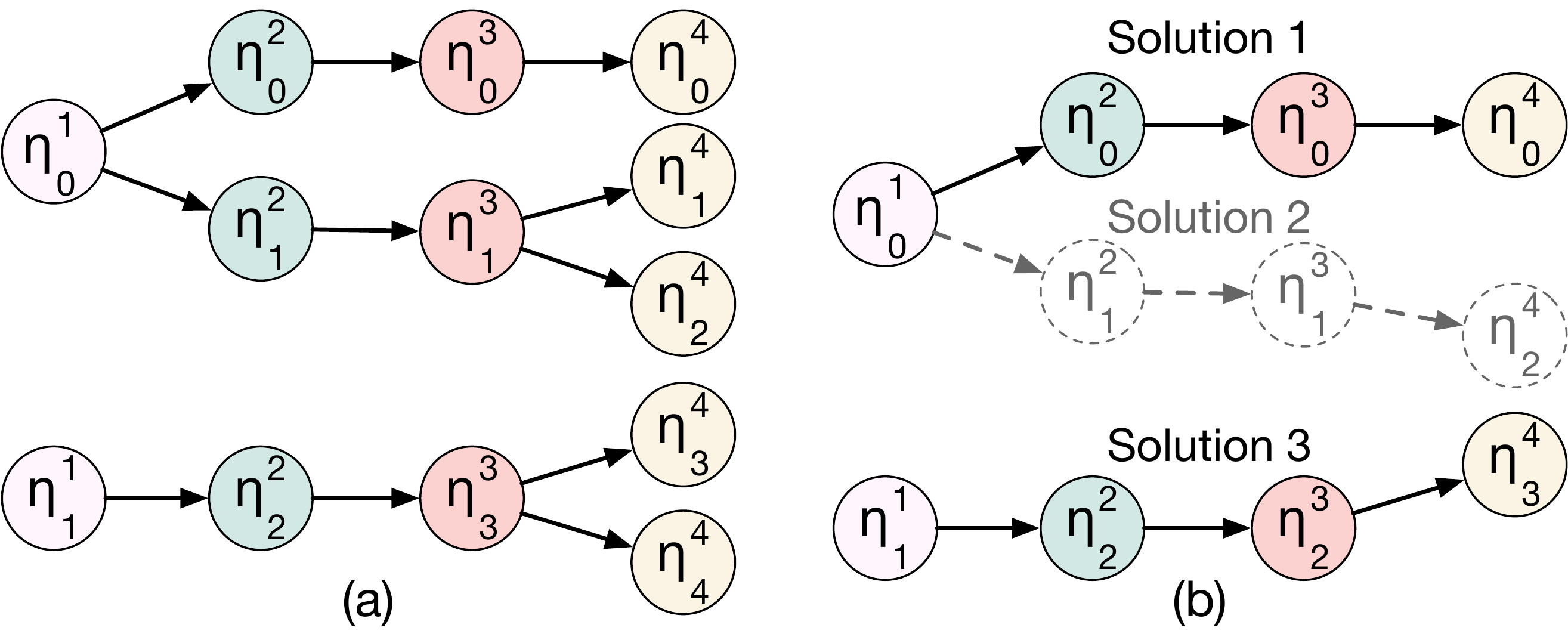}
\caption{(a) ForeSight expands the tree until the number of current solutions exceeds the maximum number allowed (four in this example). (b) It prunes the tree to only retain up to half the maximum number allowed.}
\label{fig:pathcompression}
\end{figure}

ForeSight tracks the tree size after processing each DAG layer. If the number of current solutions in the tree is less than or equal to the \textit{maximum number of solutions allowed}, it proceeds to the next DAG layer. Alternately, if the number of current solutions exceeds the maximum number of solutions allowed, ForeSight prunes the tree by evaluating the CNOT cost of each node in the most recent layer. If multiple nodes have minimum CNOT cost, ForeSight reduces the solution space to at most half the maximum number of solutions allowed. This keeps a check on the tree size while allowing it to expand by up to two solutions per node in the next layer. To maximize the diversity of the candidate pool, ForeSight only keeps one minimum-cost candidate for each mapping $\pi$. Figure~\ref{fig:pathcompression} shows an overview of Continuous Pruning where we assume a maximum of four solutions are allowed.

\begin{figure*}[htb]
\centering
  \includegraphics[width=\linewidth]{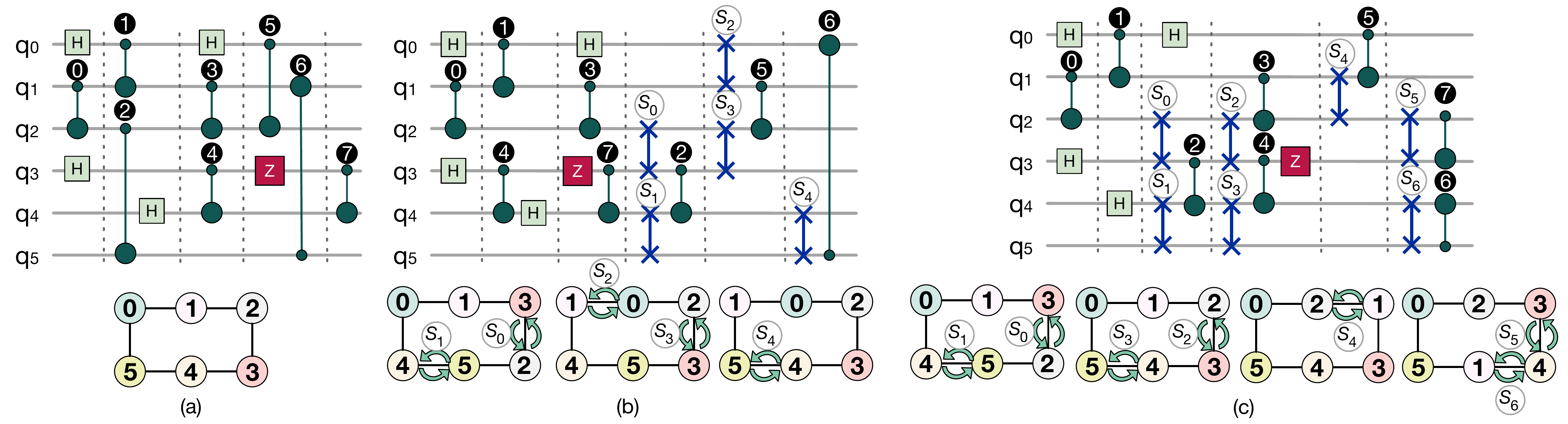}
\caption{(a) Example of a quantum circuit and its initial layout. (b) \textit{ASAP} scheduling. (c) \textit{ALAP} scheduling.}
\label{fig:asapalapscheduling}
\end{figure*}

In this example, ForeSight proceeds until layer $4$ when the number of current solutions exceeds four (equal to the number of nodes in the most recent layer of the tree). Then, it prunes the tree by evaluating the CNOT cost of nodes $\eta^4_0$ to $\eta^4_4$. Assume nodes $\eta^4_0$, $\eta^4_2$, and $\eta^4_3$ correspond to the minimum CNOT cost. As only up to four solutions are allowed, ForeSight retains two (Solutions $1$ and $3$) and discards the rest (Solution $2$) before proceeding further. Note that in our default implementation, for simplicity, the solutions to be retained are selected randomly in the event of multiple nodes corresponding to the minimum cost. This provides sufficient freedom to the retained solutions to expand in the future layers as the compilation proceeds while keeping the size of the tree tractable. ForeSight allows up to a maximum of sixty-four solutions at any time in the default implementation.

ForeSight terminates when all the DAG layers of the circuit are processed. The final schedule is generated from the solution tree by selecting the solution or branch with the minimum CNOT cost and depth at the end by tracing back up the tree from the most recent leaf node to the oldest root using the parent pointers. The overall algorithm is presented in Appendix~\ref{alg:foresight}.

\subsection{Hybrid ForeSight (ForeSight-H)}
Gate scheduling can also impact SWAP overheads. For example, IBMQ-SABRE uses \textit{As-Soon-As-Possible (ASAP)} scheduling which executes a gate whenever its dependencies are resolved~\cite{alapasap,abraham2019qiskit}. ASAP aggressively processes layers early in the circuit and requires fewer SWAPs for programs, such as \texttt{ising}~\cite{ising1925beitrag} and \texttt{qft}~\cite{coppersmith2002approximate}, where scheduling a gate resolves dependencies for gates \textit{far in future}. On the contrary, compilers (such as A*) using \textit{As-Late-As-Possible (ALAP)} scheduling, where gates are not executed until another operation can occur immediately afterward, perform poorly for such programs. For example, Figure~\ref{fig:asapalapscheduling}(a) shows a circuit and its initial layout. Figure~\ref{fig:asapalapscheduling}(b) shows ASAP scheduling. We observe that gates are scheduled as soon as their dependencies are resolved and as much as possible before SWAPs are introduced. Consequently, instructions are re-ordered (while still maintaining data dependencies) and CNOT \circled{7} is scheduled before CNOTs \circled{2}, \circled{5}, and \circled{6}. The schedule requires 5 SWAPs. Using ALAP, the same compiler introduces 7 SWAPs, as shown in Figure~\ref{fig:asapalapscheduling}(c). This is because the layout of qubits changes between the timing when the dependencies of an instruction get resolved and when it gets scheduled. For instance, here, the qubit layout changes between CNOT \circled{3} (which resolves dependencies for CNOT \circled{7}) and later when the compiler schedules CNOT \circled{7}. Our studies show that at the application-level, ALAP scheduling of A* introduces 1.6x SWAPs compared to ASAP based IBMQ-SABRE for a 16-qubit \texttt{ising} benchmark. This is consistent with prior works~\cite{li2018tackling}. Recent device-level studies show that the effectiveness of ALAP or ASAP depends on the program and device topology~\cite{smith2021error}. 

By default, ForeSight uses ALAP because (1)~it outperforms ASAP on average~\cite{smith2021error,QISKit}, and (2)~quantum programs typically do not have many parallel two-qubit gates throughout. Adding ASAP support to ForeSight frequently injects a large number of SWAP candidates to the solution tree (if there are many parallel gates in each layer) which are pruned quickly to limit the tree size, restricting their evolution in the future. The problem worsens if longer SWAP routes are retained but the cost is not amortized (mainly for programs with low depth) due to frequent pruning. Alternately, increasing the maximum number of solutions allowed worsens the compilation complexity for the average case and is not desirable. Instead, we propose \textit{Hybrid-ForeSight (ForeSight-H)} that runs ForeSight and IBMQ-SABRE concurrently and selects the schedule with minimum SWAP overheads. As IBMQ-SABRE is faster than ForeSight, ForeSight-H does not cause slowdown as the overall compilation latency is dictated by ForeSight.

\subsection{Making ForeSight Noise-Adaptive}
The error-rates on NISQ devices exhibit temporal and spatial variation.
Noise-adaptive compilers use the device error model to choose better-than-worst-case SWAPs and steer a greater number of computations to more reliable qubits and links~\cite{noiseadaptive,tannu2019not,murali2019full,murali2020architecting}. For compatibility with such policies, ForeSight makes three adjustments to the algorithm. First, it assigns weights corresponding to the CNOT error-rate to the edges of the coupling graph $G$. Second, it adds the single-qubit and measurement gates to $\mathsf{Post}$ to account for the error-rates of these operations. Finally, ForeSight chooses the output schedule based on its Expected Probability of Success (EPS)~\cite{nishio}. The EPS of a schedule is computed as the probability of successfully executing the schedule (probability that all gate and measurement operations remain error-free and no qubit decoheres) and therefore, a higher EPS is desirable. 

\color{black}

\section{Evaluation Methodology}
\label{sec:evaluation}
\subsection{Baseline Compiler}
For the baseline, we use IBM Qiskit's SABRE compiler as it is widely regarded as the state-of-the-art general-purpose compiler~\cite{QISKit,li2018tackling}. We also compare against the A* algorithm by Zulehner et al.~\cite{zulehner2018efficient} for its exhaustive search capability and Google's $t\ket{ket}$ tool-chain~\cite{googletketcompiler}. For A*, we use the QMAP tool-chain~\cite{qmap} which also includes other advanced optimizations~\cite{wille2019mapping,hillmich2021exploiting}. We use the initial mapping from SABRE, owing to its high quality, for all compilers to enable a fair comparison. Note that we particularly compare against heuristic compilers even though several recent solver-based approaches have been proposed~\cite{bhattacharjee2017depth,booth2018comparing,lye2015determining,oddi2018greedy,shafaei2013optimization,shafaei2014qubit,venturelli2017temporal,venturelli2018compiling,wille2014optimal,nannicini2021optimal} because solver-based compilers incur long latencies and therefore, have not been adopted in the industry for compiling practical applications with hundreds of qubits and operations. 

We include IBMQ's highest \textit{optimization level 3}~\cite{ibmopt3} for the baselines as well as ForeSight. This optimization performs commutative gate cancellation, re-synthesizes two-qubit unitary blocks (peep-hole optimization), uses approximate synthesis, and removes redundant resets. Its routing stage also has a larger SWAP search space. We compare with this optimization except that we disable \textit{approximation synthesis} as it is specifically for quantum volume circuits (Appendix B of~\cite{cross2019validating}) and may lead to compiled circuits that do not behave as intended for other applications~\cite{opt3issue}. The \textit{block consolidation step} combines sequences of uninterrupted gates acting on the same qubits into a unitary node and re-synthesizes them to a more compact sub-circuit.

\begin{tcolorbox}
Our baselines are industry-standard compilers from IBM and Google as well as compilers with exhaustive search capabilities. Furthermore, it also includes commutative gate cancellation, block consolidation, and re-synthesis. This establishes a fair comparison with the state-of-the-art.
\end{tcolorbox}

\subsection{Quantum Hardware Topology}
We use three topologies mentioned in Table~\ref{tab:topologies}: IBMQ Tokyo, Google Sycamore, and Rigetti Aspen9. We particularly compare against IBMQ-Tokyo as it served as the baseline for most state-of-the-art compilers~\cite{li2018tackling,tannu2019not,noiseadaptive,li2018tackling}. 

\vspace{-0.2in}
\begin{table}[htp]
\centering
\begin{small}
\caption{ Quantum Hardware Topologies}
\setlength{\tabcolsep}{0.1cm} 
\renewcommand{\arraystretch}{1.2}
\begin{tabular}{ | c | c | c |}
\hline
    Machine & Qubits & Connectivity \\
    \hline \hline
    IBMQ-Tokyo~\cite{tannu2019not,li2018tackling} & 20 & Grid with some diagonals \\
    \hline
    Google Sycamore~\cite{sycamoredatasheet,QCSup,arute2019supplementary} & 53 & Grid \\
    \hline
    Rigetti Aspen9~\cite{rigettiQPU,awsrigetti} & 32 & Octagon rings \\
    \hline
\end{tabular}
\label{tab:topologies}
\end{small}
\end{table}

\ignore{
\subsection{Enabling Other Advanced Optimizations}
IBM's optimization level 3~\cite{ibmopt3} performs commutative gate cancellation, re-synthesises two-qubit unitary blocks (peep-hole optimization), uses approximate synthesis, and removes redundant resets. Also, its routing stage has a larger SWAP search space. We enable this optimization for the baseline as well as ForeSight except that we disable the \textit{approximation synthesis pass} as it is specifically for quantum volume circuits (Appendix B of~\cite{cross2019validating}) and may lead to compiled circuits that does not behave as intended for other applications~\cite{opt3issue}. The \textit{block consolidation step} combines sequences of uninterrupted gates acting on the same qubits into a unitary node and re-synthesizes them to a potentially more optimal sub-circuit.  \ignore{Our studies show that this step favours ASAP scheduling significantly more than ALAP scheduled circuits due to the sequence of gates placed during scheduling. Hence, compilers such as SABRE greatly benefit from it. For example, enabling optimization 3 for the \texttt{adr4\_197} benchmark with A* (using ALAP) reduces SWAP overheads from 1305 to 1067 (22\% reduction), whereas SABRE (using ASAP) reduces the SWAP overheads from 1749 to 1137 (54\% reduction). As ForeSight uses ALAP scheduling by default, we modified it to use a combination of both ASAP and ALAP scheduling. Figure~\ref{fig:prelim_foresight_dynamic} show that ForeSight leads to a CNOT savings of 13\% on average and up to 22\%. The number of blocks consolidated by SABRE are 195 as opposed to 25 in ForeSight for the \texttt{adr4\_197} benchmark discussed earlier.  
}}

\subsection{Benchmarks}
We use benchmarks from the RevLib~\cite{wille2008revlib} and OpenQASM~\cite{li2020qasmbench} suites. These are widely used in several prior works including many recent studies~\cite{noiseadaptive,tannu2019not,zulehner2018efficient,li2018tackling,li2020qasmbench,siraichi2018qubit,wille2008revlib,zhang2021time,10.1145/3400302.3415620}. The benchmarks correspond to arithmetic circuits (such as hidden weighted bit, Toffoli cascades, reversible, symmetric, encoding, and decoding functions), Quantum Fourier Transform, Variational Quantum Eigensolver (VQE), and Quantum Approximate Optimization Algorithms (QAOA). Table~\ref{tab:benchmarks} summarizes some of these benchmarks but is not an exhaustive list (due to space constraints). 

\vspace{-0.2in}
\begin{table}[htp]
\centering
\begin{small}
\caption{Summary of NISQ Benchmarks}
\vspace{0.05in}
\setlength{\tabcolsep}{0.1cm} 
\renewcommand{\arraystretch}{1.2}
\begin{tabular}{ | c |c | c |}
\hline
    Benchmark  & Qubits & CNOTs \\
    \hline \hline
    
    hwb~\cite{BLSW:1999} & 4 to 6 & 107 to 2952 \\
    \hline 
    maj~\cite{FTR:2007} & 7 & 267 \\
    \hline
    
   alu~\cite{FTR:2007}& 5 to 10 & 17 to 223 \\
   \hline 
   dec~\cite{FTR:2007} & 4, 6 & 22 to 149 \\
   \hline
   
   sqrt~\cite{FTR:2007} & 7, 12 & 1314, 3087 \\
   \hline 
   ham~\cite{MDS:2005} & 3 to 15 & 9 to 3856 \\
   \hline
   
    qft~\cite{coppersmith2002approximate} & 5 to 16 & 27 to 240 \\
    \hline 
    ising~\cite{ising1925beitrag} & 10 to 16 & 90 to 150\\
    \hline
    
    vqe~\cite{teague,tomesh2022supermarq} & 6 to 8 & 933 to 4945 \\
    \hline 
    sym~\cite{FTR:2007} & 7 to 14 & 123 to 9408 \\
    \hline
    
    cmp~\cite{MDM:2005,WG:2007} & 4 to 6 & 9 to 119 \\
    \hline 
    mod~\cite{MDS:2005, WD:2009} & 5 & 13 to 78 \\
    \hline
    
    func~\cite{FTR:2007,WD:2009} & 6 to 14 & 5 to 4496 \\
    \hline
    miscellaneous~\cite{FTR:2007,MDM:2005} & 3 to 13 & 5 to 9800 \\
    \hline
    
    \ignore{
    decode~\cite{FTR:2007} & 16& 22 to 149 \\
    \hline
    sq\_root~\cite{FTR:2007} & 16 & 869 \\
    \hline
    ham~\cite{MDS:2005} & 16 & 11, 149\\
    \hline
    miller~\cite{WG:2007} & 16 & \\
    \hline
    qft~\cite{coppersmith2002approximate} & 10, 16 & 90, 240\\
    \hline
    ising~\cite{ising1925beitrag} & 10, 13, 16 & 90, 120, 150\\
    \hline
    sym~\cite{FTR:2007} \\
    \hline
    vqe~\cite{teague} & 6, 7, 8 & \\
    \hline}
\end{tabular}
\label{tab:benchmarks}
\end{small}
\end{table}

\ignore{
\begin{table}[htp]
\vspace{-0.05in}
\centering
\begin{small}
\caption{\color{blue}Summary of NISQ Benchmarks}
\setlength{\tabcolsep}{0.1cm} 
\renewcommand{\arraystretch}{1.2}
\begin{tabular}{ | c |c | c | c | }
\hline
    Benchmark & Acronym & Application & CNOT \\
    \hline \hline
    Hidden Weighted Bit & \texttt{hwb} & BDD Models~\cite{BLSW:1999} & \\
    \hline
    ESOP & \texttt{alu} &  Toffoli Cascade~\cite{FTR:2007} & \\
    \hline
    \multirow{3}{*}{Encoding Functions} & \texttt{decode} & Binary Decoder & \\
    \cline{2-4}
    & \texttt{graycode} & Graycode Decoder & \\
    \cline{2-4}
    & \texttt{ham} & Hamming Decoder & \\
    \hline
    \multirow{3}{*}{Miscellaneous} & \multirow{3}{*}{-} &  Arithmetic Functions & \\\cline{3-4}
     & & Reversible Functions & \\\cline{3-4}
     & & Symmetric Functions & \\
    \hline
    Fourier Transform & \texttt{qft} & Shor's Algorithm & \\
    \hline
    
\end{tabular}
\label{tab:benchmarks}
\end{small}
\end{table}
}

\color{black}

\subsection{Figure-of-Merit}
We quantify the performance of a compiler based on its ability to reduce SWAP overheads in terms of the number of CNOTs (Equation~\ref{eq:swapoverheads}) and depth. This metric has been widely used in several prior works~\cite{zulehner2018efficient,li2018tackling,siraichi2018qubit,gerard2021affinemap,tan2021gateabsorb,liu2021qucloud,nannicini2021optimal}. \textit{Thus, lower SWAP overheads and circuit depth are desirable. }
\begin{equation}
\begin{split}
\textrm{SWAP Overheads} =  \textrm{CNOTs in compiled program} - \\
\textrm{CNOTs in actual program}
\end{split}\label{eq:swapoverheads}
\end{equation}

\subsection{Evaluations on Noisy Simulator}
To study the impact of noise-adaptive ForeSight, we model the error rates of Google Sycamore~\cite{sycamoredatasheet} and use a depolarizing noise channel\cite{depolarizing}. To account for crosstalk, we consider the \textit{simultaneous} error-rates that are derived from cross-entropy benchmarking~\cite{boixo2018characterizing}. We measure the \textit{Fidelity} of a program~\cite{patel2021robust,jigsaw,adapt,ash2020experimental,sarovar2020detecting} by computing the Total Variation Distance~\cite{TVD} between the noise-free ($p^{\textrm{ideal}}$) and noisy ($p^{\textrm{noisy}}$) output distributions, as shown in Equation~\eqref{eq:fidelity}. \textit{Fidelity $\in [0,1] $  and a higher Fidelity is desirable. }\begin{equation}
\textrm{Fidelity}(p^{\textrm{ideal}},p^{\textrm{noisy}}) = 1- \sum_{i=1}^{k}\mid \mid p^{\textrm{ideal}}_i - p^{\textrm{noisy}}_i \mid \mid )
\label{eq:fidelity}
\end{equation}


\section{Final Evaluations}
\subsection{Impact on SWAP Overheads}
\begin{figure}[htb]
  \includegraphics[width=\columnwidth]{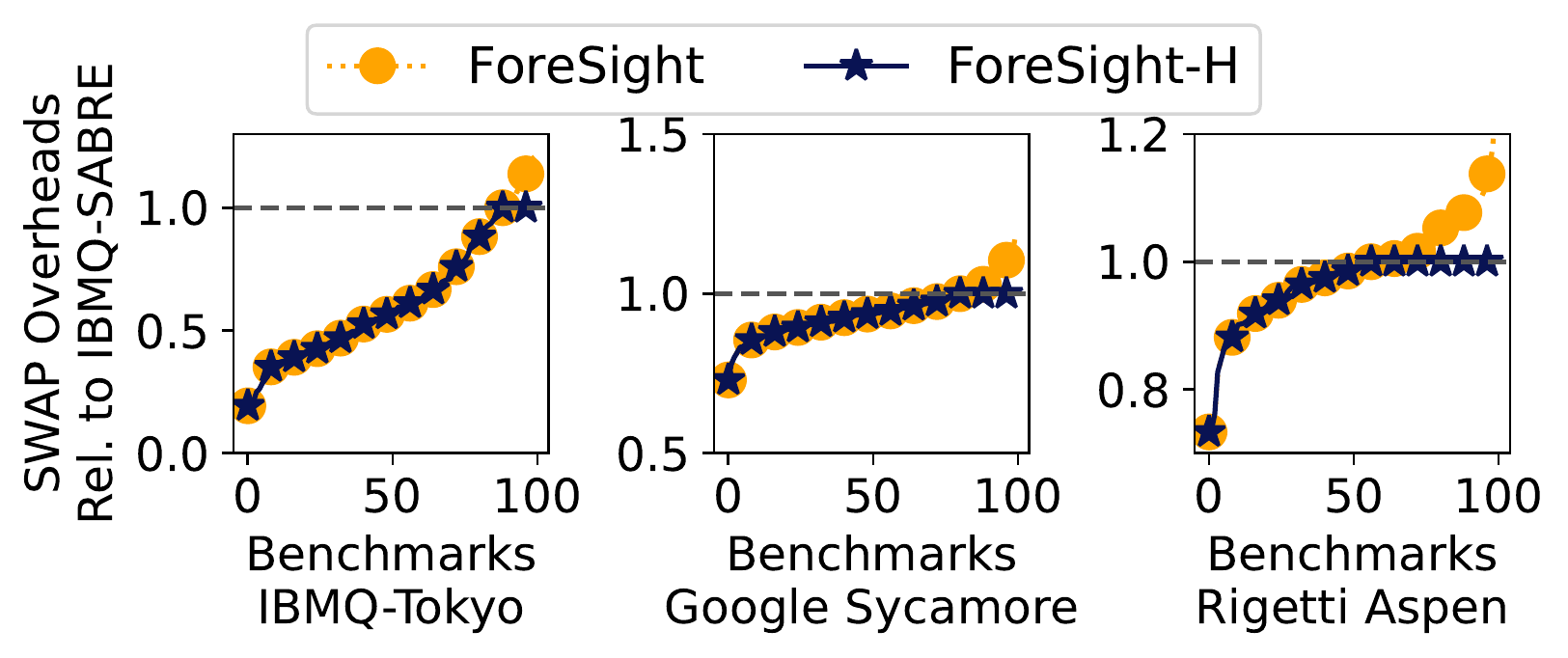}
\caption{Relative SWAP Overheads of ForeSight}
\label{fig:resultsummarywithoutopt3}
\end{figure}

\noindent \textbf{Results Summary:} Figure~\ref{fig:resultsummarywithoutopt3} shows the SWAP overheads of ForeSight relative to IBMQ-SABRE with optimization level 3 enabled. ForeSight leads to CNOT savings of 50\%, 14\%, and 6\% on average for IBMQ-Tokyo, Google Sycamore, and Rigetti Aspen respectively. On these devices, ForeSight leads to CNOT savings up-to 86\%, 25\%, and 27\% respectively compared to IBMQ-SABRE. 
ForeSight's CNOT savings are 43\%, 6\%, and 2\% on average and up-to 81\%, 27\%, and 27\% for these devices respectively when optimization level 3 is enabled. \textit{Thus, ForeSight is effective even in the presence of other advanced optimizations.}

Table~\ref{tab:swapaverages} shows the CNOT savings for different program sizes without and with optimization level 3. The effectiveness of ForeSight increases with program size. On Rigetti Aspen, ForeSight has limited ability to explore multiple candidates and the utility of evaluating longer SWAP routes diminish. 

\vspace{-0.1in}
\begin{table}[htp]
\centering
\begin{small}
\caption{Average and Best-Case CNOT Savings (in percentage) based on Program Size (Higher is Better)}

\setlength{\tabcolsep}{0.2cm} 
\renewcommand{\arraystretch}{1.2}
\begin{tabular}{ | c | c |c||c|c|| c| c| }
\hline
    
    Program & \multicolumn{2}{c||}{IBMQ-Tokyo} & \multicolumn{2}{c||}{Google Sycamore} & \multicolumn{2}{c|}{Rigetti Aspen} \\
    \cline{2-7}
    CNOTs & Avg & Max & Avg & Max & Avg & Max \\
    \hline \hline
     \multicolumn{7}{|c|}{Without Optimization level 3} \\
     \hline
     0 - 2K & 48.6 & 85.7 & 13.0 & 24.5 & 5.92 & 25.9 \\
\hline
2K - 5K & 58.7 & 67.9 & 16.8 & 20.0 & 8.16 & 13.8 \\
\hline
5K - 10K & 57.1 & 76.2 & 15.2 & 16.9 & 10.2 & 27.1 \\
\hline
    \hline
     \multicolumn{7}{|c|}{With Optimization level 3} \\
    \hline \hline
0 - 2K & 40.8 & 80.6 & 5.9 & 27.0 & 1.28 & 26.0 \\
\hline
2K - 5K & 52.1 & 61.8 & 6.61 & 12.1 & 3.56 & 7.36 \\
\hline
5K - 10K & 49.6 & 71.5 & 5.97 & 10.5 & 6.19 & 26.6 \\
\hline

\end{tabular}
\label{tab:swapaverages}
\end{small}\vspace{-0.05in}
\end{table}

\begin{figure}[tb]
  \includegraphics[width=\columnwidth]{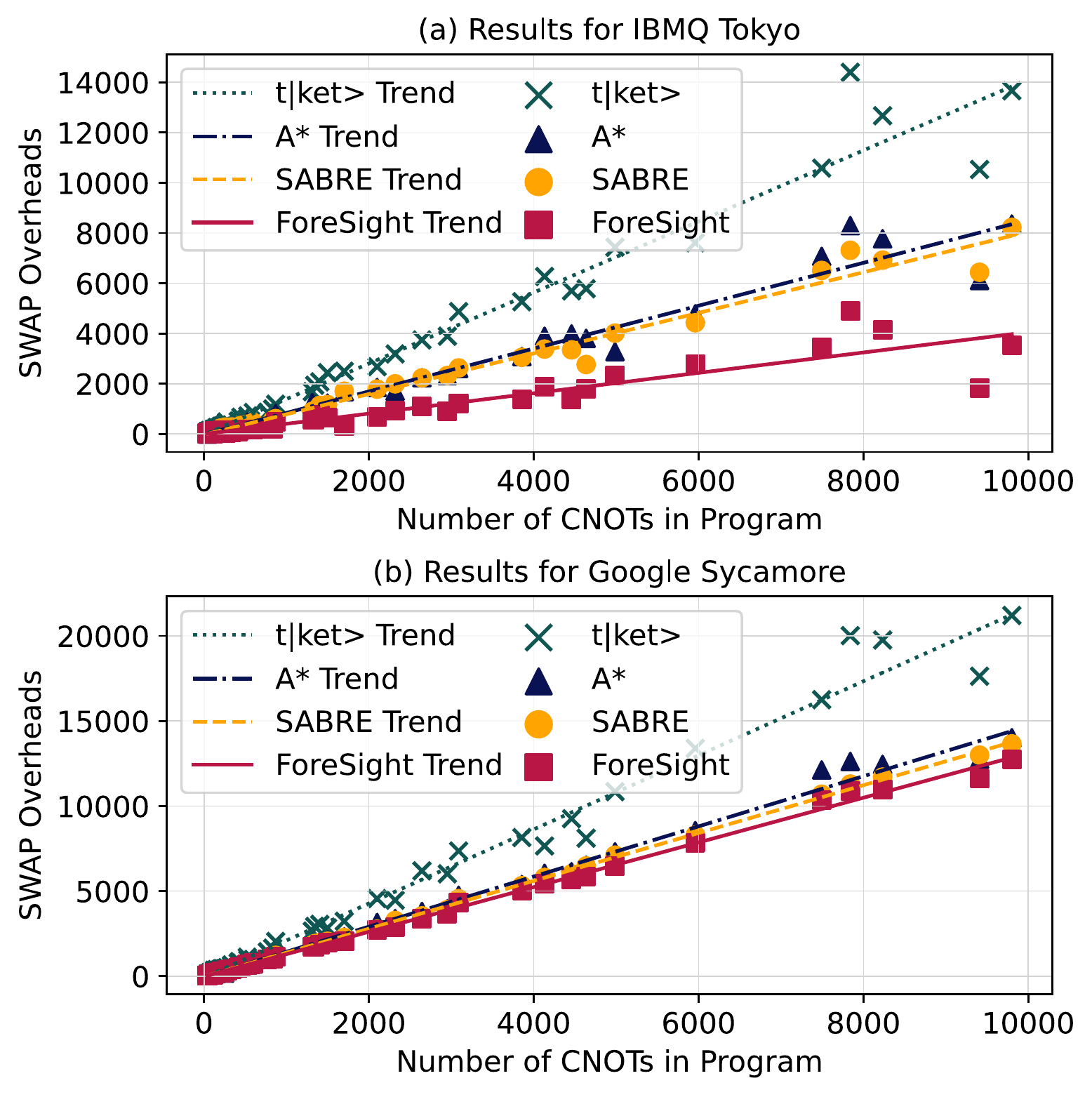}
\caption{SWAP overheads (with optimization 3) vs. program size on (a) IBMQ-Tokyo (b) Google Sycamore.}
\label{fig:absolutecnots}
\end{figure}

Figure~\ref{fig:absolutecnots} compares the SWAP overheads on IBMQ-Tokyo and Google Sycamore. ForeSight outperforms $t\ket{ket}$, IBMQ-SABRE, and A$^*$ for circuits with hundreds of CNOTs, the scale needed for practical  problems~\cite{bian2019quantum,nam2020ground,basso2021quantum,lolur2021benchmarking,kim2021universal,guerreschi2019qaoa}. For small circuits, ForeSight performs comparable to IBMQ-SABRE as the low gate count and depth reduce the utility of assessing multiple solutions and slightly longer SWAP routes. Note that A* standalone significantly outperforms SABRE, but SABRE reduces the performance gap with optimization level 3 because of the difference in scheduling. The ASAP scheduling of SABRE allows a greater number of blocks to be consolidated compared to A* which uses ALAP. Table~\ref{tab:compilercomparisonrawnumbers} compares SWAP overheads for some selected benchmarks.

\vspace{-0.15in}
\begin{table}[!htb]
\begin{center}
\centering
\footnotesize
\caption{Comparison of Compilers Across  Devices}
\vspace{-0.05in}
\setlength{\tabcolsep}{1.1mm} 
\renewcommand{\arraystretch}{1.1}
\label{tab:compilercomparisonrawnumbers}
{\footnotesize
\begin{tabular}{ |c|c||c|c|c|c|} 

\hline
Benchmark & 
\multirow{2}{*}{Machine} & \multicolumn{4}{c|}{SWAP Overheads (\textit{with} optimization 3)} \\ \cline{3-6}
(CNOTs) & & $t|ket\rangle$ & SABRE & A$^*$ & ForeSight \\
\hline
\multirow{2}{*}{wim\_266} & Tokyo & 512 & 267 & 329 & \textbf{159} \\
\cline{2-6}
 & Sycamore & 855 & 548 & 577 & \textbf{532} \\
\cline{2-6}
(427) & Aspen & 907 & \textbf{596} & 710 & 610 \\
\hline

\multirow{2}{*}{misex1\_241} & Tokyo & 2677 & 1785 & 1853 & \textbf{681} \\
\cline{2-6}
 & Sycamore & 4559 & 2760 & 3160 & \textbf{2734} \\
\cline{2-6}
(2100) & Aspen & 3721 & 2993 & 3531 & \textbf{2896} \\
\hline

\multirow{2}{*}{vqe\_n7} & Tokyo & 446 & 230 & 243 & \textbf{-15} \\
\cline{2-6}
 & Sycamore & 188 & \textbf{32} & 52 & 43 \\
\cline{2-6}
(2306) & Aspen & 1064 & 321 & 693 & \textbf{119} \\
\hline

\multirow{2}{*}{hwb6\_56} & Tokyo & 3891 & 2326 & 2323 & \textbf{909} \\
\cline{2-6}
 & Sycamore & 6015 & 3908 & 3965 & \textbf{3658} \\
\cline{2-6}
(2952) & Aspen & 5126 & 4103 & 4214 & \textbf{3801} \\
\hline

\multirow{2}{*}{sqn\_258} & Tokyo & 5698 & 3348 & 3987 & \textbf{1386} \\
\cline{2-6}
 & Sycamore & 9256 & 6017 & 6139 & \textbf{5666} \\
\cline{2-6}
(4459) & Aspen & 8592 & 6471 & 7751 & \textbf{6253} \\
\hline

\multirow{2}{*}{root\_255} & Tokyo & 10584 & 6510 & 7072 & \textbf{3456} \\
\cline{2-6}
 & Sycamore & 16249 & 10691 & 12124 & \textbf{10349} \\
\cline{2-6}
(7493) & Aspen & 18607 & 10691 & 14064 & \textbf{10658} \\
\hline

\multirow{2}{*}{life\_238} & Tokyo & 13665 & 8225 & 8358 & \textbf{3517} \\
\cline{2-6}
 & Sycamore & 21199 & 13654 & 14018 & \textbf{12729} \\
\cline{2-6}
(9800) & Aspen & 20117 & 18944 & 22570 & \textbf{13894} \\
\hline

\end{tabular}}
\end{center}
\vspace{-0.1in}
   \begin{tablenotes}
   \footnotesize
   \item $^*$\textit{Negative SWAP overheads indicate reduction in gate count post block consolidation and gate cancellations.}
    \end{tablenotes}
\end{table}

\vspace{0.05in}
\noindent \textbf{When does ForeSight not outperform IBMQ-SABRE?}
 We observe some cases where IBMQ-SABRE outperforms ForeSight. For example, IBMQ-SABRE outperforms ForeSight for 8 out of 100 benchmarks on IBMQ-Tokyo. These correspond to small circuits with 38 to 124 CNOTs or are programs that benefit from the ASAP scheduling of SABRE, such as qft and ising benchmarks. This latter observation is consistent with prior works~\cite{li2018tackling}. Also, there is a difference in the SWAP insertion policy between IBMQ-SABRE and ForeSight. Unlike IBMQ-SABRE which introduces one SWAP at a time, ForeSight adds an entire routing path of SWAPs. Thus, the compilers explore different search spaces. 

\textit{ForeSight-H prevents performance degradation} by running ForeSight and IBMQ-SABRE concurrently and selecting the best schedule between them without degrading compilation time. ForeSight-H benefits from the diversity of (1)~scheduling and (2)~swap insertion policies of both compilers.

\subsection{Impact on Circuit Depth}
Table~\ref{tab:depthnumbers} shows the circuit depth from ForeSight relative to IBMQ-SABRE and A*. ForeSight reduces the depth to 0.93x on average and up-to 0.67x compared to the baseline.\ignore{\textit{ Note that ForeSight  schedules have both lower SWAP overheads and depth and thus, has higher EPS compared to baseline. }
}

\begin{table}[!htb]
\begin{center}
\begin{small}
\caption{Circuit depth relative to IBMQ-SABRE and A*}
\setlength{\tabcolsep}{1.2mm} 
\renewcommand{\arraystretch}{1.2}
\label{tab:depthnumbers}
{\footnotesize
\begin{tabular}{ |c|c|c||c|c|} 
\hline
\multirow{2}{*}{Machine} & \multicolumn{2}{c||}{IBMQ-SABRE} &  \multicolumn{2}{c|}{A* Routing Algorithm} \\ 
\cline{2-5}
& Average & Best-Case & Average & Best-Case\\
\hline
\hline

IBMQ Tokyo & 0.83 & 0.67 & 0.85 & 0.71 \\
\hline
Google Sycamore & 0.96 & 0.78 & 0.98 & 0.74 \\
\hline
Rigetti Aspen & 0.97 & 0.84 & 0.98 & 0.74 \\
\hline
\end{tabular}}
\end{small}
\end{center}
\end{table}

\noindent \textbf{Comparison with Compiler that Optimizes for Depth}: TOQM trades off SWAP cost for circuit depth~\cite{zhang2021time}. Table~\ref{tab:toqmcomparison} shows the ratio of SWAP overheads, circuit depth, and Expected Probability of Success (EPS) of ForeSight to TOQM on Sycamore. EPS quantifies the impact of the trade-off between SWAPs and depth. ForeSight outperforms TOQM if EPS ratio is more than 1. The EPS ratio increases with program size because the mean CNOT error-rate and coherence time on Sycamore is 1\% and 15 $\mu$seconds respectively and thus, the extra SWAPs reduce the EPS of TOQM schedule. 
\begin{table}[htb]
\begin{center}
\begin{small}
\caption{Ratio of SWAP Overheads, Depth, and EPS}
\setlength{\tabcolsep}{1.2mm} 
\renewcommand{\arraystretch}{1.2}
\label{tab:toqmcomparison}
{\footnotesize
\begin{tabular}{ |c|c|c|c|c|} 
\hline
Program & CNOTs & SWAP Ratio & Depth Ratio & EPS Ratio\\
\hline
\hline

rev\_syn\_n8 & 97 & 0.79 & 1.08 & 1.07\\
\hline
wim\_266 & 427 & 0.80 & 1.03 & 2.10 \\
\hline 
sqrt\_n7 & 3089 & 0.68 & 1.08 & $3.89 \times 10^{16}$ \\
\hline
\end{tabular}}
\end{small}
\end{center}
\end{table}

\begin{figure}[!htb]
\centering
\vspace{-0.1in}
  \includegraphics[width=\columnwidth]{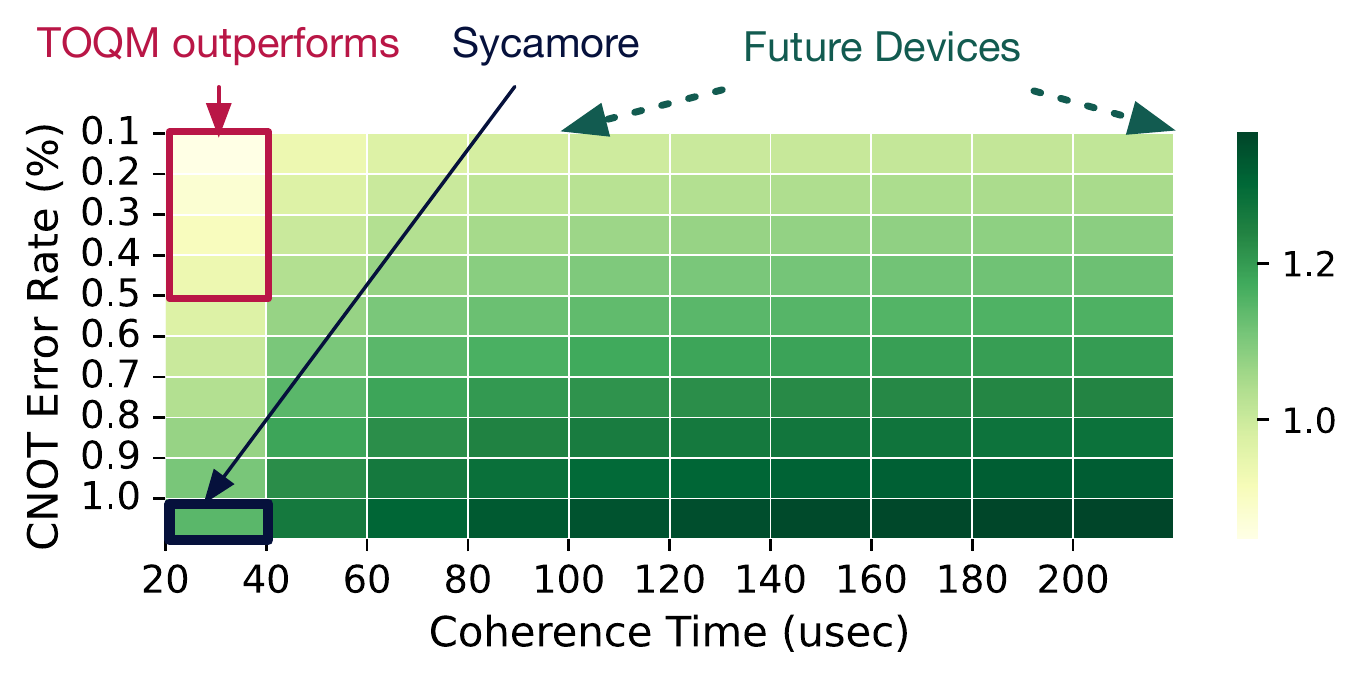}
\caption{EPS Ratio of ForeSight to TOQM}
\label{fig:epsratioheatmap}
\end{figure}

To understand the trade-off, we compute the EPS ratio for various coherence times and CNOT error-rates for 8-qubit \texttt{rev\_syn\_n8} benchmark. We assume single-qubit and two-qubit gate latencies as 25ns and 32ns respectively~\cite{sycamoredatasheet}. ForeSight's schedule has 218 CNOTs and a depth of 8521ns. TOQM's schedule has 251 CNOTs and a depth of 8024ns. Figure~\ref{fig:epsratioheatmap} shows that ForeSight generates schedules with higher EPS than TOQM except for very low coherence times and CNOT error-rates. This is unlikely as we expect both CNOT error-rates and coherence times to improve in future.

\subsection{Impact on Probability of Success}
To assess the overall performance, we show the EPS ratio of ForeSight to IBMQ-SABRE for some benchmarks in Table~\ref{tab:epsr}. We use the error-rates of Sycamore as well as two optimistic models, M1 and M2. For M1, we assume 100 $\mu$s coherence time and 0.5\% CNOT and measurement error-rates. For M2, these are 250 $\mu$s and  0.1\% respectively.  We observe that even though the EPS ratio decreases when error-rates improve by an order of magnitude (expected as CNOT errors go down), it is still more than 1. The EPS ratio of \texttt{vqe\_n8} is lower than other programs with more CNOTs because of gate cancellations. On Sycamore, the EPS ratio is high for large benchmarks due to very high CNOT error-rates (1\%). 

\begin{table}[!htb]
\begin{center}
\begin{small}
\caption{EPS Ratio of ForeSight to IBMQ-SABRE}
\setlength{\tabcolsep}{1.2mm} 
\renewcommand{\arraystretch}{1.2}
\label{tab:epsr}
{\footnotesize
\begin{tabular}{ |c|c|c|c|c|} 
\hline
Program & CNOTs & Sycamore & Model M1 & Model M2 \\
\hline
\hline

hwb4\_49 & 105 & 2.73 & 1.27 & 1.07 \\
\hline
con1\_216 & 415 & 3.32 & 1.49 & 1.10 \\
\hline
vqe\_n8 & 4945 & 4.95 & 1.58 & 1.13 \\
\hline
cycle10\_2\_110 & 2644 & 8955 & 7.10 & 1.86 \\
\hline
sym9\_148 & 9408 & $2.9 \times 10^{11}$ & 6634 & 8.50 \\
\hline
\end{tabular}}
\end{small}
\end{center}
\end{table}

\subsection{Scalability: Memory-Time Complexity}
NISQ compiler complexity depends on program as well as the device characteristics. If $T_g$ and $C_{\textrm{depth}}$ are the number of gates and circuit depth respectively, ForeSight processes $\frac{T_g}{C_{\textrm{depth}}}$ gates in each DAG layer on average. Assume each gate requires SWAP(s). To find a SWAP candidate, ForeSight folds all paths from the distance matrix. The number of paths for a candidate depends on the mean device connectivity ($\mu_G$), number of physical qubits ($N$), and constraint relaxation factor ($\delta$) and scales $O(\mu_G^\delta N^2)$. Assuming the solution tree is at maximum capacity, there are $\frac{\mathrm{S_{max}}}{2}$ nodes in the most recent layer of the tree, where $S\mathrm{_{max}}$ is the maximum number of solutions allowed. In the worst-case, each node searches for a SWAP candidate for $\frac{T_g}{C_{\textrm{depth}}}$ gates and repeats for $C_{\textrm{depth}}$ DAG layers. Thus, the complexity of ForeSight scales $O(\frac{S\mathrm{_{max}}}{2}\times{C_{\textrm{depth}}}\times \frac{T_g}{C_{\textrm{depth}}}\mu_G^\delta N^2$). Simplifying, we get Equation~\eqref{eq:complexity}.

\begin{equation}
O(\frac{S\mathrm{_{max}}}{2}\times T_g\mu_G^\delta N^2)
\label{eq:complexity}
\end{equation}

Thus, the complexity of ForeSight scales quadratic with the number of physical qubits and linear with gates in a program. We implement ForeSight in Python and evaluate on an Intel Xeon Gold core with 16 GB memory and operating at 3.5 GHz. Table~\ref{tab:timecomplexity} compares the time complexity. Note that the A* compiler is in C++, unlike SABRE and ForeSight. In general, A* search scales exponentially~\cite{martelli1977complexity}, and the original implementation~\cite{zulehner2018efficient} required more than 32 GB for programs with just 16-20 qubits~\cite{li2018tackling}. The A* compiler has improved in the last three years~\cite{wille2019mapping,hillmich2021exploiting} but, we were still unable to run programs beyond 100 qubits. On the contrary, ForeSight's average memory usage is within 200 MB for our evaluations. Even if the complexity of A* improves, ForeSight still outperforms it in terms of SWAP overheads.

\begin{table}[htb]
\begin{center}
\begin{small}
\caption{Compilation Latency in seconds}
\setlength{\tabcolsep}{1.2mm} 
\renewcommand{\arraystretch}{1.2}
\label{tab:timecomplexity}
{\footnotesize
\begin{tabular}{ |c|c|c||c|c||c|c|} 
\hline
\multirow{2}{*}{Machine} & \multicolumn{2}{c||}{A* (C++)} &  \multicolumn{2}{c||}{SABRE (Python)} &  \multicolumn{2}{c|}{ForeSight (Python)} \\ 
\cline{2-7}
& Mean & Max & Mean & Max & Mean & Max\\
\hline
\hline

Tokyo  & 2.15 & 19.1 & 0.90 & 8.30 & 48.5 & 783 \\
\hline
Sycamore & 2.29 & 20.3 & 1.15 & 10.6 & 300 & 3888 \\
\hline
Aspen & 2.39 & 22.7 & 1.14 & 12.2 & 194 & 2981 \\
\hline
\end{tabular}}
\end{small}
\end{center}
\end{table}


We also study the complexity of up-to 500-qubit Bernstein Vazirani programs~\cite{bernsteinvazirani} encoding ``all ones" keys.  Table~\ref{tab:qaoatimecomplexity} shows the memory and time complexity on a 25x20 grid. 
\vspace{-0.15in}
\begin{table}[htb]
\begin{center}
\begin{small}
\caption{Memory (MB) and Time (sec) Complexity}
\setlength{\tabcolsep}{1.1mm} 
\renewcommand{\arraystretch}{1.1}
\label{tab:qaoatimecomplexity}
{\footnotesize
\begin{tabular}{ |c|c|c|c||c|c|c|} 
\hline
Program & \multicolumn{3}{c||}{IBMQ-SABRE} &  \multicolumn{3}{c|}{ForeSight} \\ 
\cline{2-7}
Qubits & CNOTs & Memory & Time &  CNOTs & Memory & Time \\
\hline
\hline

\ignore{
100 &  381  &  3.97  &  0.30  &  264  &  22.0  &  92 \\
\hline
200 &  741  &  3.97  &  0.57  &  558  &  65.2  &  168 \\
\hline
500 &  2727  &  6.52  &  2.00  &  1884  &  153  &  413 \\
\hline
}
100 &  426  &  1.54  &  0.16  &  285  &  17.0  &  66 \\
\hline
200 &  888  &  2.54  &  0.32  &  603  &  36.7  &  130 \\
\hline
500 &  2742  &  5.77  &  1.01  &  2055  &  83.1  &  321 \\
\hline

\end{tabular}}
\end{small}
\end{center}\vspace{-0.1 in}
\end{table}

\subsection{Comparison of SWAP Insertion}
\begin{figure}[!htb]
  \includegraphics[width=\columnwidth]{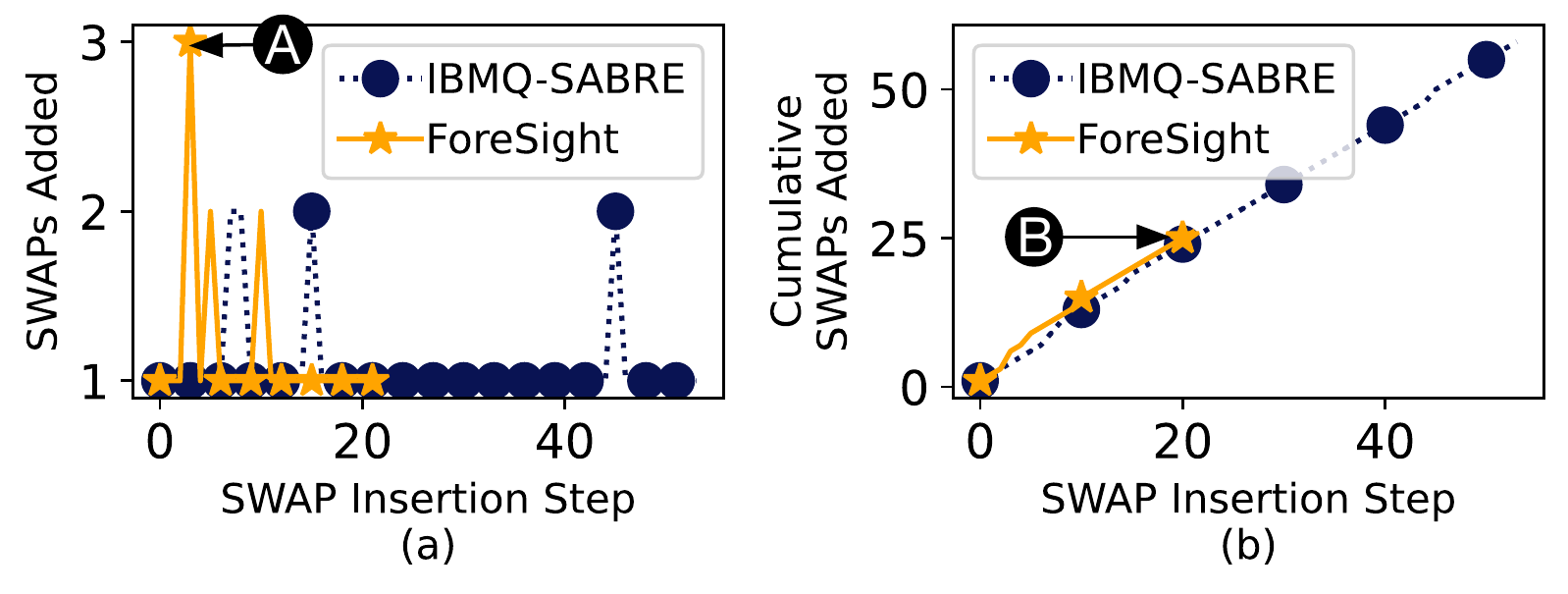}

\caption{SWAP insertion comparison for \texttt{vqe\_n8}.}
\label{fig:vqeswapanalysis}
\end{figure}

Figure~\ref{fig:vqeswapanalysis} shows the individual and cumulative number of SWAPs introduced by IBMQ-SABRE and ForeSight for a 8-qubit VQE benchmark on Sycamore. ForeSight chooses three SWAPs at Step-\circled{A} unlike one SWAP in IBMQ-SABRE. This results in a reduced number of SWAP insertion steps at application-level and fewer SWAPs overall. For example, IBMQ-SABRE introduces 58 SWAPs, whereas ForeSight completes with fewer steps at \circled{B} and costs only 26 SWAPs. Thus, unlike existing compilers, ForeSight reduces SWAPs globally at the application-level.

\subsection{Complexity vs. Quality of Solutions}
\begin{figure}[!htb]
  \includegraphics[width=\columnwidth]{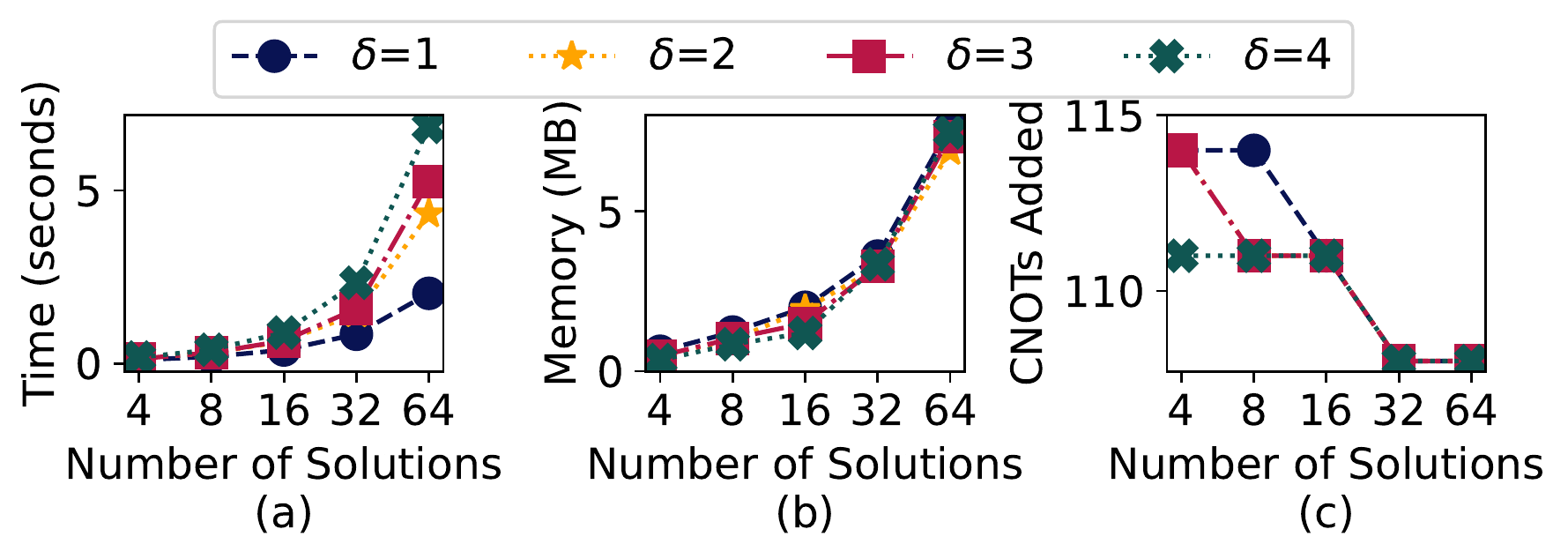}
  \caption{(a) Time (in seconds), (b) Memory (in MB), and (c) SWAP overheads in a 50-qubit BV benchmark on Google Sycamore for different values of $\delta$ and $S\mathrm{_{max}}$.}
\label{fig:bv50tradeoff}
\end{figure}

Figure~\ref{fig:bv50tradeoff} shows the trade-offs in memory-time complexity and quality of solutions for a 50-qubit BV benchmark on Google Sycamore topology. The solution search space of ForeSight and therefore, the memory and time complexity increases with the increase in maximum number of solutions ($S\mathrm{_{max}}$) and constraint relaxation factor ($\delta$). Increasing $\delta$ but limiting $S\mathrm{_{max}}$ to a small value may reduce performance as good solutions get frequently discarded during the pruning step. Also, increasing $S\mathrm{_{max}}$ increases the complexity significantly but leads to diminishing returns beyond a certain point because many of the additional solutions in the tree are weak. Our experiments with several benchmarks show that $S\mathrm{_{max}} = 64$ is sufficient on average and achieves a sweet spot in terms of both complexity as well as quality of solutions. The impact of increasing $\delta$ diminishes beyond $\delta=2$ as the additional cost of introducing longer SWAP routes in the solution space does not get amortized. 

\ignore{
Figure~\ref{fig:sensitivitytobranches} shows the SWAP overheads relative to IBMQ-SABRE for different values of $S\mathrm{_{max}}$ on Sycamore (for $\delta = 2$). Our experiments show that $S\mathrm{_{max}} = 64$ is sufficient on average. 

\begin{figure}[!htb]
  \includegraphics[width=\columnwidth]{./Figures/sensitivity_to_smax.pdf}
\caption{Relative SWAP overheads for different $S\mathrm{_{max}}$.}
\label{fig:sensitivitytobranches}
\end{figure}
}

\subsection{ForeSight vs. More Iterations of SABRE}
\begin{figure}[htb]
  \includegraphics[width=\columnwidth]{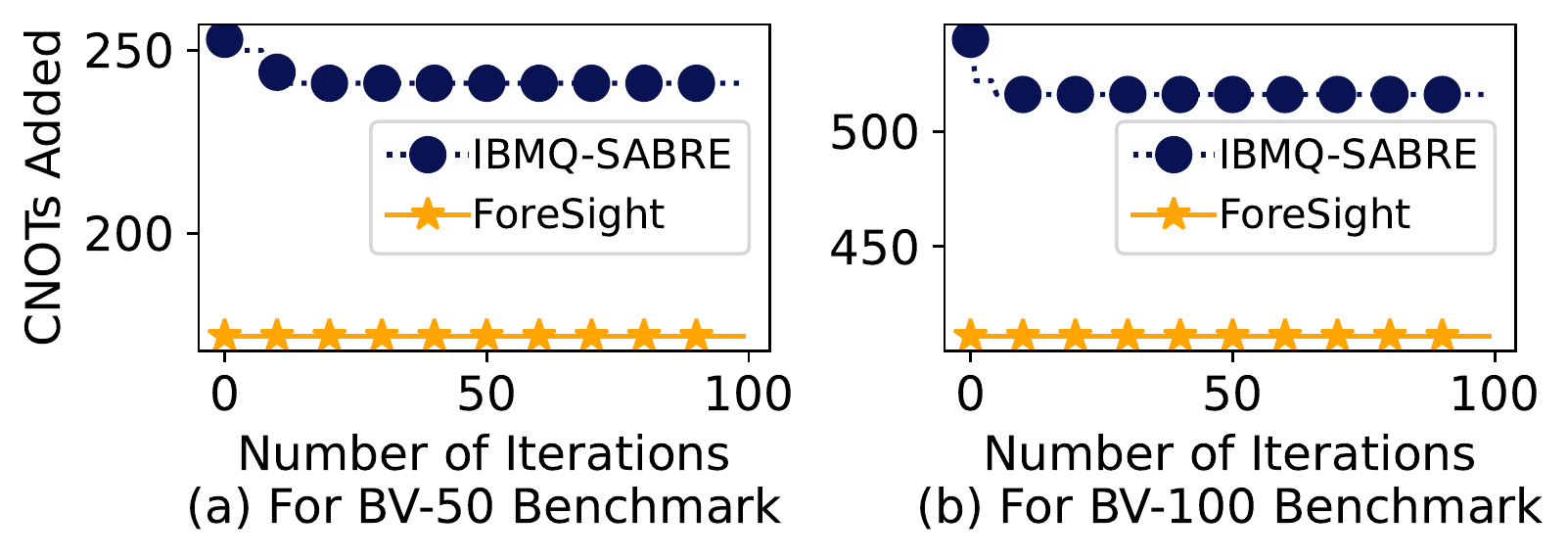}
\caption{SWAP overheads with increasing iterations of both IBMQ-SABRE and ForeSight for two different BV circuits.}
\label{fig:bvconvergence}
\vspace{0.2in}
\end{figure}

Figure~\ref{fig:bvconvergence} shows the SWAP overheads with increasing iterations. The solutions from ForeSight within one iteration outperform SABRE's solutions even after a hundred iterations. Thus, although ForeSight is slower than SABRE, spending the time difference on an increased number of iterations for SABRE does not match the performance of ForeSight. 

\subsection{Results for Noise-Adaptive ForeSight}
\begin{figure}[!htb]
\centering
  \includegraphics[width=\columnwidth]{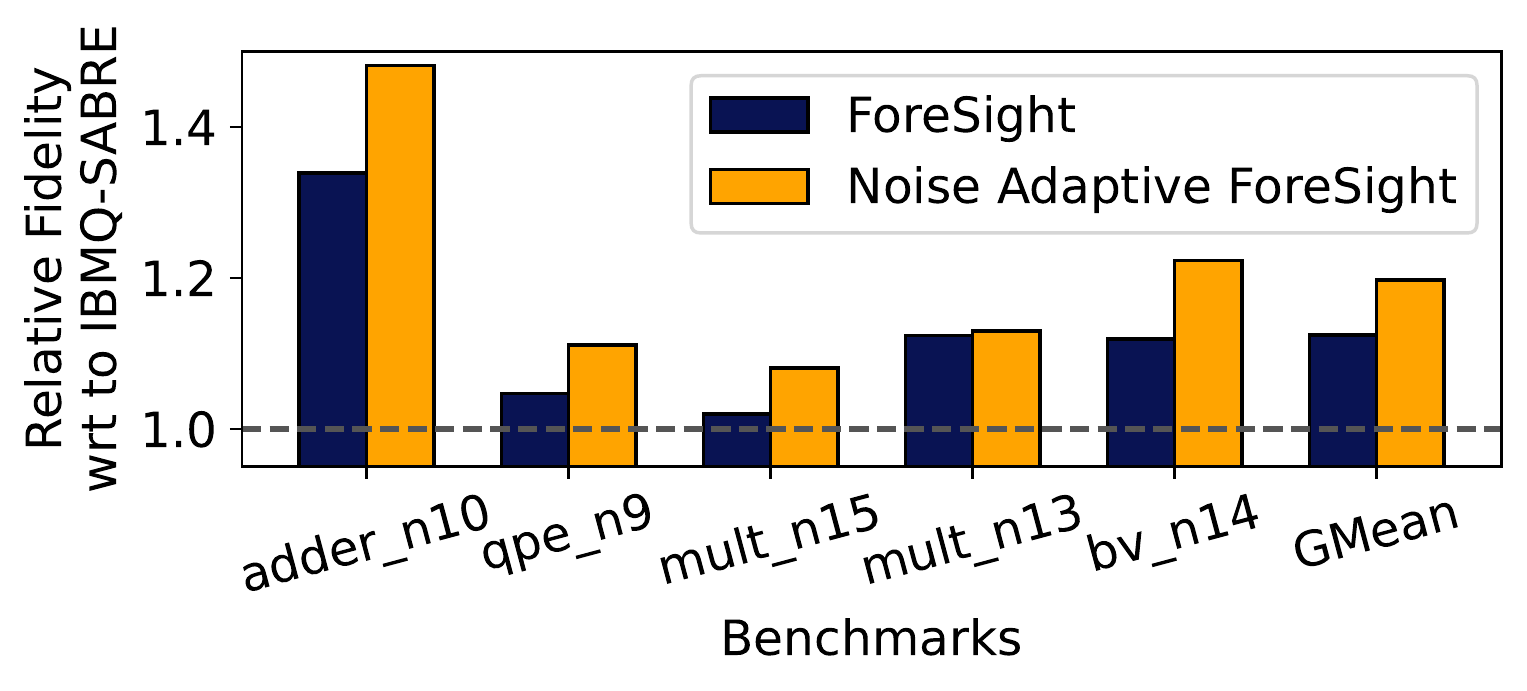}
\caption{Fidelity of ForeSight and Noise-Adaptive ForeSight relative to IBMQ-SABRE.}
\label{fig:fidelityimpact}
\vspace{0.2in}
\end{figure}

Figure~\ref{fig:fidelityimpact} shows that noise-adaptive ForeSight improves Fidelity by 1.17x on average and by up-to 1.41x compared to the baseline on a noisy simulator using the error-rates of Google Sycamore. Our studies show that in certain programs, noise-adaptive ForeSight improves Fidelity despite using more CNOTs by selecting SWAPs with higher fidelity. 

\color{black}

\color{black}

\color{black}
\section{Conclusion}
Near-term quantum computers have restricted connectivity between qubits and SWAP operations introduced by compilers to overcome this limitation present a serious bottleneck in running large quantum programs with high fidelity.
In this paper, we propose ForeSight, a compiler that minimizes SWAP overheads in NISQ programs. \textit{Firstly}, ForeSight evaluates slightly longer qubit movement paths for the current operations that can potentially lead to fewer SWAPs while scheduling future operations. \textit{Secondly}, ForeSight evaluates multiple SWAP candidates and delays SWAP selection, thereby preventing early convergence on sub-optimal SWAPs.  \textit{Finally}, to keep the search tractable, ForeSight continuously adapts the solution space by adding newer candidates and eliminating weaker ones. Our evaluations with a hundred representative benchmarks and three different hardware topologies show that ForeSight reduces SWAP overheads by 17\% on average and by up-to 81\%, compared to the baseline.

\newpage
\section*{APPENDIX}
\label{sec:foresight}
The ForeSight routing algorithm is summarized in Algorithm~\ref{alg:foresight}.
\setlength{\textfloatsep}{0pt}
\SetAlgoNoLine
{
\begin{algorithm}[htb]
\caption{ForeSight Routing Algorithm}
\label{alg:foresight}
\SetKwInput{KwInput}{Input}                
\SetKwInput{KwOutput}{Output}              
\DontPrintSemicolon
  
  \KwInput{\textrm{\textbf{(1)} Program} \textrm{\textbf{(2)} Device distance matrix:} $D[][]$ and routing capacity $\mu_G$ \textrm{\textbf{(3)} Constraint relaxation $\delta$}}
  \KwOutput{\textrm{Machine Compatible Schedule}}

  \SetKwFunction{FBupdate}{Preprocessing}
  \SetKwFunction{Fcreatepool}{Get\_Pool}
  \SetKwFunction{Fschedule}{Compile\_Program}
 
  \ignore{
  \SetKwProg{Fn}{Function}{:}{\KwRet}
  \Fn{\FBupdate{$G,\delta$}}{


        $\mu_G \leftarrow \frac{\mid E \mid}{\mid V \mid}$\; 
        \For{each $i,j$ in $N$:}{
                $D[i][j]$ = [ ]; $ [P_{i \rightarrow j}] \leftarrow$ paths from $Q_i$ \textrm{to }$Q_j$\; 
                \For{$P$ in $[P_{i \rightarrow j}]$:}{
                \If{ $d(P^{\textrm{min}}_{i \rightarrow j}) \leq d(P) \leq d(P^{\textrm{min}}_{i \rightarrow j})+\delta$:}{
                    $D[i][j]$.append($P$)\;}
                }
                
        }
        \KwRet $D,\mu_G$\;
  }}
  \SetKwProg{Fn}{Function}{:}{}
  \Fn{\Fcreatepool{$D$,$F$,$\mathsf{Post},\pi,\mu_G$}}{

        $\mathrm{All\_Paths} \leftarrow [P_{i \rightarrow j} \in D[\pi(g)], \forall g \in F]$\;
       $\mathrm{Swap\_Candidates} \leftarrow [\mathrm{fold}(P), \forall P\in \mathrm{All\_Paths}]$ \;
       $\mathrm{Scores} \leftarrow $ Compute $H_{total}$ for $\mathrm{Swap\_Candidates}$\;  
       \ignore{
       \For{path in candidate\_list:}{
       SWAP\_candidate = [folded$\_$path $ \forall g \in F]$\;
       Cost = $H_{total}($SWAP$\_$candidate,$\mathsf{Post},\mu_G)$ \; 
       Scores.append([SWAP\_candidate,Cost])
       }}
       \KwRet $\mathrm{Swap\_Candidates}$ with min($\mathrm{Scores}$)\;


  } 

  \SetKwProg{Fn}{Function}{:}{}
  \Fn{\Fschedule{Program,$D$,$\mu_G$}}{
  
    Initialize $\pi(), F, \mathsf{Post}$, and solution tree $T$\; 
    \While{$|F| > 0$}{
        \For{$\mathrm{Parent} \in \mathrm{leaves}(T)$}{
            $\mathrm{Pool} = $ \Fcreatepool{$D,F,\mathsf{Post},\pi,\mu_G$}\;
            
            \For{$\mathrm{Swap\_Candidate} \in \mathrm{Pool}$}{
                Add solution $S$ as child of $\mathrm{Parent}$\;
                Update $\pi()$ and compute CNOT cost\;
            }
            
            \If{$N_\mathrm{solutions}(T) > \mathrm{Max\_Solutions}$}{
                Prune $T$\;
            }
            $F \leftarrow \mathrm{next}(F)$ and update $\mathsf{Post}$\;
        }
        \KwRet Schedule: $S$ in $T$ with min CNOT cost
    }
\ignore{
       Solution tree $T$ = empty; Initialize $\pi(),F,\mathsf{Post}$; \;       
       \While{$F$ is not empty:}{
       
       \For{parent in most recent leaves of $T$:}{
        $Pool$ = \Fcreatepool{$D$,$F$,$\mathsf{Post},\pi,\mu_G$}\;

       \For{SWAP\_candidate in Pool:}{
       Add $S$: $parent \rightarrow child$; 
       Update $\pi()$\;
       Compute $\mathsf{CNOT}$ Cost of $child$\;
       }
       
       }
       \If{\#Current Solutions of $T$ $>max_{\textrm{solutions}}$:}
       {Prune $T$\;}
       $F \leftarrow$ next layer; $\mathsf{Post} \leftarrow $ Future $Layers$ of $F$\;
    } 
    Schedule $\leftarrow$ $S$ of $T$ with $min(\mathsf{CNOT}$ Cost)\;
    \KwRet Schedule\;
    
}
    
  }

\end{algorithm}}

\hfill 
\section*{ACKNOWLEDGEMENT}
We thank the reviewers of ISCA-2022 for their comments and feedback. We thank Nicolas Delfosse and Yunong Shi for helpful discussions and comments. We also thank Moumita Dey, Sanjay Kariyappa, Narges Alavisamani, and Ramin Ayanzadeh for their editorial suggestions. Poulami Das was funded by the Microsoft Research PhD Fellowship. 

\hfill \break

\bibliographystyle{plain}
\bibliography{references}

\end{document}